%% file: main.tex
\documentclass[11pt]{article}

\usepackage[margin=1in]{geometry}
\usepackage[T1]{fontenc}
\usepackage{epstopdf}
\usepackage{enumitem}
\setlist{nosep}

\usepackage{url}
\usepackage{graphicx}
\usepackage{float}
\usepackage{amsmath}
\usepackage{amssymb}
\usepackage{amsthm}
\usepackage{bm}

\usepackage{tabularx, booktabs, ragged2e}
\usepackage[round]{natbib}
\newcolumntype{Y}{>{\RaggedRight\arraybackslash}X}

\graphicspath{{figures/}}

\theoremstyle{plain}
\newtheorem{theorem}{Theorem}[section]

\newtheorem{proposition}[theorem]{Proposition}
\theoremstyle{definition}

\theoremstyle{remark}

\newcommand{\skewt}{\mathcal{S}t}                 % skewed Student-t
\newcommand{\dd}{\mathrm{d}}
\newcommand{\E}{\mathbb{E}}
\newcommand{\Prob}{\mathbb{P}}
\newcommand{\Normal}{\mathcal{N}}
\newcommand{\given}{\,|\,}
\newcommand{\DRF}{\mathrm{DRF}}
\newcommand{\Black}{\mathrm{Black}}
\newcommand{\cdfmc}{\widehat{F}_N}                % empirical (MC) CDF
\newcommand{\Lmin}{\mathcal{L}_{\min}}
\newcommand{\Lcdf}{\mathcal{L}_{\mathrm{cdf}}}
\newcommand{\netparams}{\psi}                     % neural-network weights
\newcommand{\modelparams}{\Theta}                 % volatility-model parameters

\newcounter{paperpart}
\newcommand\partdivider[1]{%
  \par\addvspace{1.1\baselineskip}\refstepcounter{paperpart}%
  {\centering\large\bfseries Part \Roman{paperpart}\quad #1\par}%
  \addcontentsline{toc}{part}{Part \Roman{paperpart}\quad #1}%
  \addvspace{0.5\baselineskip}\nopagebreak}

\begin{document}

\title{Fast, Reliable, and Error-Bounded Option Pricing with Pretrained
Neural Networks: A GJR--GARCH Study}

\author{Thijs van den Berg\thanks{The author thanks Paul Wilmott for
valuable feedback and discussions.}\\
Simu Labs \\ \texttt{thijs@simu.ai}}

\date{}
\maketitle

\begin{abstract}
Many models in quantitative finance have no closed-form option prices and rely on
slow, noisy Monte Carlo simulation; neural surrogates restore speed but offer no
error guarantees. We present a general recipe for surrogates that are fast, with
bounded and verifiable error, applicable to any simulation-based density model. A
Mixture Density Network maps parameters and maturity to the terminal return
density as a Gaussian mixture, so prices, implied volatilities, and Greeks follow
in closed form as an arbitrage-free mixture of lognormals, with a CDF-matching
loss aligned to pricing error. A distribution-free Monte Carlo noise floor,
$\sqrt{1/(6N)}$, quantifies the best accuracy achievable at a given simulation
budget and decomposes the out-of-sample error into four controllable terms. We
demonstrate the method on GJR--GARCH, where the surrogate reaches an
out-of-sample CDF error of $1.4\times10^{-4}$, within $10\%$ of the noise floor,
and prices each option in a few microseconds on a single CPU core, or under a
microsecond on a GPU.
\end{abstract}



\input{sections/01_introduction}
\input{sections/02_related_work}
\partdivider{A Framework for Error-Bounded Surrogates}
\input{sections/03_densities_to_prices}
\input{sections/04_mixture_surrogate}
\input{sections/05_cdf_loss}
\input{sections/06_noise_floor}
\input{sections/07_error_anatomy}
\input{sections/08_design_rules}
\partdivider{Application to GJR--GARCH}
\input{sections/09_gjr_garch_model}
\input{sections/10_training_data}
\input{sections/11_capacity_optimisation}
\partdivider{Evaluation}
\input{sections/12_results}

\input{sections/13_pricing_behaviour}
\input{sections/14_discussion}

\section*{Declaration of interest statement}
The author reports no conflict of interest.

\section*{Data availability statement}
The training dataset is publicly available on the Hugging Face Hub at
\texttt{simu-ai/garch\_densities}. All experiment code (training, error analysis
and figure generation) is released alongside this manuscript.

\bibliographystyle{plainnat}
\bibliography{references}

%----------------------------------------------------------
\appendix
\partdivider{Appendices}
\input{sections/appendix_a_gjr_garch}
\input{sections/appendix_b_data_training}
\input{sections/appendix_c_notation}

\end{document}

%% file: sections/01_introduction.tex
\section{Introduction}

Option pricing in realistic volatility models often requires simulation. The
GJR--GARCH process \citep{glosten1993relation} is a prime example: it captures
asymmetry, fat tails, and volatility clustering, yet has no closed-form
pricing formula. In practice Monte Carlo simulation \citep{Glasserman2004} is the
only general approach, but its computational cost makes large-scale calibration
and risk management impractical, and the resulting prices still carry sampling
noise. These limitations are most acute when a pricing surface must be revalued
repeatedly across many parameter configurations.

Neural networks offer a way to bypass simulation, but most existing approaches
are either narrowly tuned to specific payoffs or opaque, with
limited guarantees on accuracy. For financial applications speed alone is not
sufficient: reliability and verifiability are equally critical.

In this work we introduce a pretrained neural network that serves as a drop-in
surrogate for Monte Carlo under the GJR--GARCH model. Rather than regressing on
option prices, we learn the \emph{terminal return density} implied by the model.
Once this density is known, prices and Greeks follow analytically by integrating
Black-type payoffs against it, so a single predicted density delivers the prices
of all strikes at a given maturity, as well as risk metrics such as
Value-at-Risk, without further simulation or numerical integration. The network
thus acts as a learned, Fokker--Planck--type \emph{forward operator}: it returns
the finite-horizon density implied by the model without explicit time stepping.

\paragraph{Contributions}
Our contributions are both methodological and architectural: a general,
model-agnostic recipe for turning any Monte Carlo density simulator into a fast
surrogate with quantified, verifiable error, realised as a forward density
operator and demonstrated end-to-end on GJR--GARCH.

\begin{enumerate}[leftmargin=*]
  \item \textbf{A forward density operator.} A Mixture Density Network (MDN) maps
  parameters and maturity $(\modelparams, T)$ to the parameters of a Gaussian
  mixture for the terminal return density (equivalently, a mixture of lognormals
  in price); option prices and most Greeks then follow in closed form,
  arbitrage-free by construction (Sections~\ref{sec:densities}
  and~\ref{sec:surrogate}).
  \item \textbf{A loss aligned with pricing.} Training minimises the
  root-mean-square deviation between the predicted and Monte Carlo cumulative
  distribution functions (CDFs) over a uniform quantile grid; for
  lognormal-mixture models this CDF deviation bounds the option-price error
  across strikes (Section~\ref{sec:loss}).
  \item \textbf{A closed-form Monte Carlo noise floor.} The sampling noise of the
  training targets, $\sqrt{q(1-q)/N}$, integrates to a distribution-free accuracy
  floor $\Lmin=\sqrt{1/(6N)}$ that ties the achievable error to the simulation
  budget alone, independent of the model being learned (Section~\ref{sec:floor}).
  \item \textbf{An error anatomy and design rules.} The floor anchors a split of
  the out-of-sample error into four approximately independent terms (label noise,
  network capacity, parameter coverage, and optimiser temperature), each with a
  control and an observable, turning a target precision into explicit design
  rules and a principled stopping point
  (Sections~\ref{sec:anatomy} and~\ref{sec:design}).
  \item \textbf{A demonstration on GJR--GARCH.} We apply the recipe to a model
  with no closed-form pricer, removing its scale and shift symmetries via a
  dimensionless reduced form and covering the parameter space with Sobol
  low-discrepancy sampling so that test errors match training errors across the
  admissible region (Sections~\ref{sec:model} and~\ref{sec:data}). The trained
  surrogate reaches an out-of-sample CDF error within $10\%$ of the floor and prices
  each option in a few microseconds on a single CPU core, or under a microsecond on a
  GPU (Section~\ref{sec:results}).
\end{enumerate}

Together these elements make the surrogate fast, reliable, and verifiable: an
error-certified replacement for Monte Carlo, trained to the accuracy limit of
simulation itself. The microsecond-scale pricing this enables brings GARCH-class
models within reach of settings where simulation has been prohibitive (real-time
market-making and quoting, intraday portfolio risk and VaR, and large-scale
calibration), and the recipe transfers to any simulation-based density model.

\paragraph{How this paper is organised}
The paper has three parts. The first develops the method in general terms,
without tying it to any one model. Section~\ref{sec:densities} shows that an
option price follows directly from the distribution of future returns, so the
only thing a model must get right is that distribution, and that a mixture of
normals prices exactly, as a weighted sum of closed-form Black prices.
Section~\ref{sec:surrogate} represents that distribution with a Mixture Density
Network whose outputs price in closed form.
Section~\ref{sec:loss} introduces the training objective, which matches the
predicted and simulated distributions, and shows that doing so keeps the pricing
error small. Section~\ref{sec:floor} explains why there is a hard accuracy limit:
because the training targets come from simulation, no surrogate can do better than
the simulation noise itself, which is $\sqrt{1/(6N)}$ for $N$ paths.
Section~\ref{sec:anatomy} then splits the test error into four sources, each with
one thing that controls it and one signal that reveals it, and
Section~\ref{sec:design} turns that split into explicit design rules for reaching a
target precision. Together they set up the rest of the paper.

The second part applies the method to a concrete example, GJR--GARCH, a model with
no closed-form pricer. Section~\ref{sec:model} defines the model, rewrites it in
a dimensionless reduced form that removes its redundant scale and shift parameters
before training, and standardises the surrogate's target onto a common scale
across maturities. Section~\ref{sec:data} covers the training data: how the return
distributions are simulated, and how the parameter space is sampled evenly with a
Sobol design. Section~\ref{sec:errors} then shows how to push the two
training-related errors down and measures the constants the design rules need.

The third part reports the results. Section~\ref{sec:results} measures how close
the trained surrogate gets to the limit, shows which parameter regions are
hardest, identifies what controls the error, and maps the trade-off between
accuracy, model size, and speed. Section~\ref{sec:behaviour} illustrates the
option-price surfaces the model produces, and Section~\ref{sec:discussion}
concludes.

%% file: sections/02_related_work.tex
\section{Related Work}
\label{sec:related}
The use of neural networks for option pricing dates to the early 1990s, when
\citet{Hutchinson1994} trained a multilayer perceptron to learn the
Black--Scholes pricing map directly from observed S\&P~500 futures option
prices, building on \citet{Malliaris1993}. The survey of \citet{Ruf2020}
catalogues over a hundred of the papers that followed and organises the
landscape.

\paragraph{Model-free vs.\ model-based surrogates}
The literature splits into two families. \emph{Model-free} approaches, in the
original Hutchinson--Lo--Poggio paradigm, train on listed market prices and
benchmark against the market itself, acting as a non-parametric smoother of the
observed surface and inheriting its noise and arbitrage violations.
\emph{Model-based} approaches instead emulate the theoretical prices of a
parametric model, a fast surrogate for an underlying numerical scheme. The
present work is model-based: the ``truth'' we measure against is the GJR--GARCH
model itself, not the market.

\paragraph{Model-based surrogates}
Within model-based surrogates, the key distinction is the training signal. When
the model already has a (near-)closed-form pricer, the network fits a
deterministic map against exact targets, with no sampling noise:
\citet{Hernandez2017} emulated Heston calibration orders of magnitude faster than
direct optimisation, \citet{McGhee2018} the analogue for SABR, and
\citet{LiuOosterleeBohte2019} for Black--Scholes implied volatilities. The harder
case, and ours, is models without closed-form prices, where the targets
themselves come from simulation and carry its noise. \citet{Horvath2019} and
\citet{Bayer2018} pioneered this for rough volatility, and
\citet{Liu2019Framework} cast it as a calibration framework; \citet{Itkin2019}
catalogues the recurring pitfalls, limited extrapolation and accuracy plateaus,
many of them direct consequences of the Monte Carlo noise floor of
Section~\ref{sec:floor}. Pricing under GARCH dynamics is itself simulation-based
except in special cases, from the Monte Carlo valuation of \citet{Duan1995} to
the closed-form affine-GARCH of \citet{HestonNandi2000}; the general GJR
specification we use has no closed-form pricer and must be simulated. Closest to
our model choice, \citet{Wang2009GJR} used a hybrid GJR--GARCH/NN model for
stock-index option prediction, but model-free and market-fit rather than as a
simulation surrogate.

\paragraph{Density- and structure-aware networks}
A separate strand, closer to our output representation, predicts the
\emph{distribution} of returns rather than option prices.
\citet{Schittenkopf2001} use a mixture density network to extract risk-neutral
densities from option prices, the inverse of our problem, in the same
architectural family. Structural priors that reduce input dimension also predate
our reparameterisation: \citet{GarciaGencay2000} exploit the homogeneity of the
Black--Scholes pricer in moneyness, and \citet{Dugas2009} bake monotonicity and
convexity into the architecture. Arbitrage-free parameterisations of the
implied-volatility surface itself were studied by \citet{GatheralJacquier2014}
(SVI).

\paragraph{Positioning}
We sit at the intersection of these strands: a model-based surrogate for a
simulation-only volatility model, restricted to European options, whose network
outputs a return distribution. What distinguishes the contribution is its
treatment of accuracy. Where earlier simulation surrogates meet the Monte Carlo
plateau as a pitfall to be managed \citep{Itkin2019}, we recast it as a
closed-form, distribution-free \emph{design target} (Section~\ref{sec:floor}) and
give a model-agnostic recipe for reaching it. Instantiated on GJR--GARCH, this
yields what is, to our knowledge, the first real-time pricer for the general
(non-affine) specification: fast enough to bring the model into the latency- and
compute-bound applications where simulation has been impractical.

%% file: sections/03_densities_to_prices.tex
\section{Density-Based Option Pricing}
\label{sec:densities}

Like most realistic volatility models, GJR--GARCH does not allow options to be
priced by perfect replication: with only the underlying and a bond to trade, a
nonlinear payoff cannot be exactly reproduced, so no-arbitrage alone leaves some
pricing freedom. We adopt the standard resolution for this setting
\citep{Duan1995} and value by expectation: the underlying carries a risk-neutral
drift (the risk-free rate, since the option is hedged with the underlying), and
an option's value is the discounted expected payoff under the model-implied
terminal \emph{return} distribution. 

Once the terminal return density $p(x)$ is known, the price of any payoff
$g(F_T)$ follows from
\begin{equation}
  C = e^{-rT}\int g\!\big(F_0 e^{x}\big)\,p(x)\,\dd x,
  \label{eq:expectation_price}
\end{equation}
where $F_0$ is the forward and $r$ the risk-free rate. It is convenient to model
\emph{returns}: let $X_T$ be the log-return over horizon $T$, so the terminal
forward is $F_T = F_0 e^{X_T}$. Whatever shape we choose for $p$, the mean of
$X_T$ is pinned by the no-arbitrage forward condition
\begin{equation}
  \E[F_T] = F_0 e^{rT},
  \label{eq:forward}
\end{equation}
so only the \emph{shape} of $p$ (skew, tails, multimodality) remains to be
modelled.

\paragraph{The Black family as a density}

Seen this way, the classical Black models are just the Gaussian special case.
Black-Scholes (equities), Black
(futures), and Garman--Kohlhagen (FX) all assume
$X_T\sim\Normal(-\tfrac12\sigma^2 T,\,\sigma^2 T)$, yielding
\begin{align}
  C   &= e^{-rT}\big(F_0\,\Phi(d_1) - K\,\Phi(d_2)\big), \label{eq:black}\\
  d_1 &= \frac{\ln(F_0/K) + \tfrac12\sigma^2 T}{\sigma\sqrt{T}}, \qquad
  d_2  = d_1 - \sigma\sqrt{T},\nonumber
\end{align}
where the asset class enters only through the choice of forward $F_0$
(Appendix~\ref{app:data}). The key observation is that $\Phi(d_2)$ is the
risk-neutral probability of finishing in the money,

\begin{equation}
  \Phi(d_2) = \Prob^{\mathbb{Q}}\!\big(X_T > \ln(K/F_0)\big)
\end{equation}

Option valuation depends only on the return CDF; the other inputs
$F_0, K, r, \sigma$ just set dimensional scales. Replacing $\Phi$ by any other
CDF generalises pricing to a different return distribution, which is exactly why
we make CDF accuracy the core learning objective (Section~\ref{sec:loss}).

\begin{figure}[t]
  \centering
  \includegraphics[width=\textwidth]{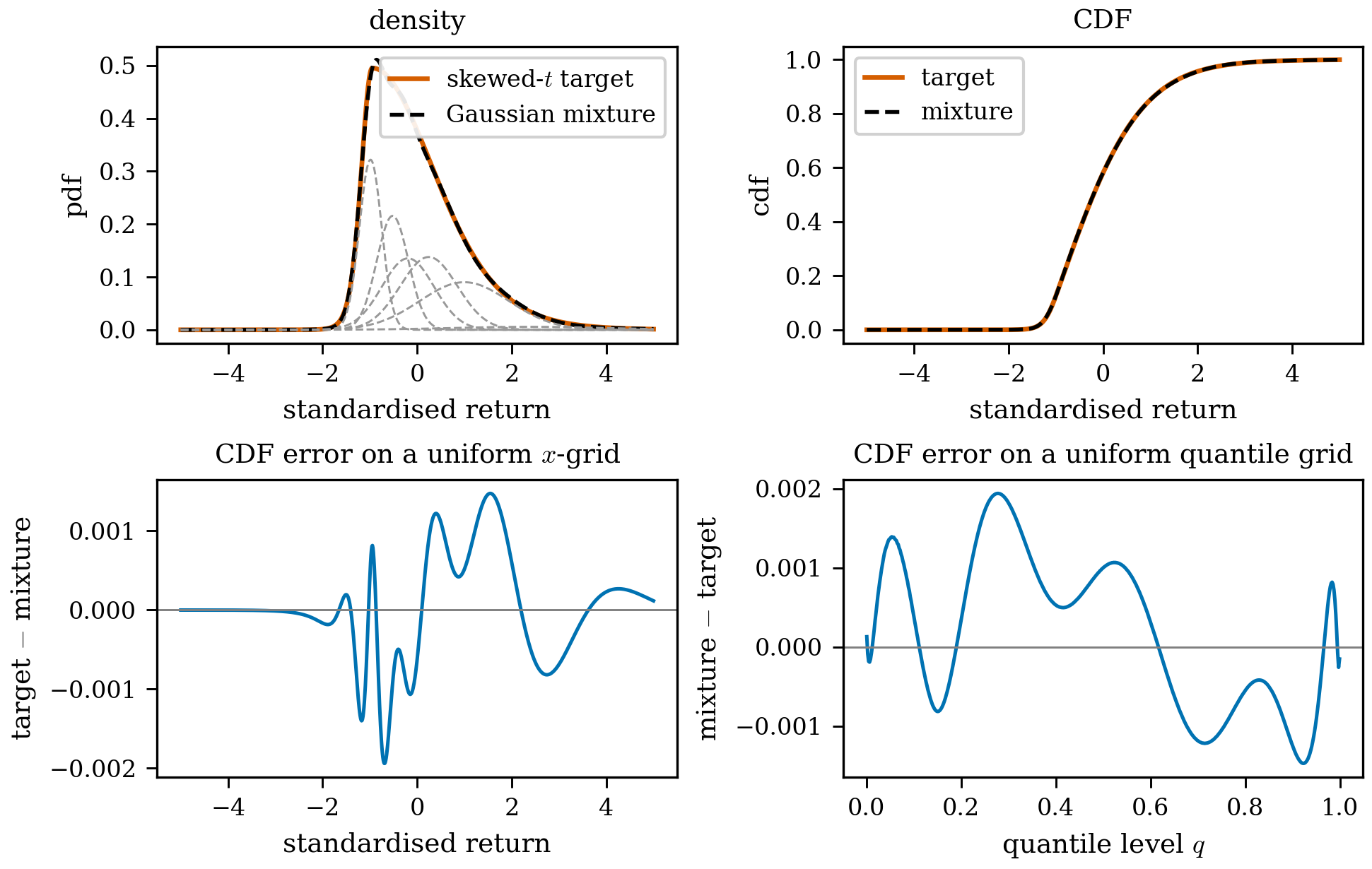}
  \caption{\textbf{A few-component Gaussian mixture represents a return density
    accurately.} The mixture (black) matches a skewed-$t$ density and CDF (red),
    with components dashed. The lower panels compare CDF error in $x$-space
    (left), where the far-left tail contributes almost nothing to the RMSE, with
    error sampled uniformly in quantile space (right), which gives equal weight to
    all probability mass and motivates our CDF-over-quantiles loss.}
  \label{fig:gmm_pdf_cdf}
\end{figure}

\paragraph{Gaussian mixtures and ``a sum of Blacks''}
To represent non-Gaussian shapes we use a Gaussian Mixture Model (GMM),
\begin{equation}
  p(x) = \sum_{i=1}^{M} w_i\,\Normal(x\given\mu_i,\sigma_i^2),
  \qquad w_i\ge 0,\ \sum_i w_i = 1.
  \label{eq:gmm_density}
\end{equation}
A GMM can approximate any smooth density to arbitrary accuracy, and in practice
only a few components capture skew, heavy tails, or multimodal structure
(Fig.~\ref{fig:gmm_pdf_cdf}). Each component is normal in log-returns, which is
exactly the lognormal price model Black solves in closed form. The total option
value is therefore a weighted sum of component Black prices,
\begin{align}
  F_i &= F_0\,e^{\mu_i' + \tfrac12\sigma_i^2}, \qquad
  \hat\sigma_i = \sigma_i\sqrt{T}, \nonumber\\
  C(K) &= \sum_{i=1}^{M} w_i\,\Black\!\big(F_i, K, \hat\sigma_i, r, T\big),
  \label{eq:mixture_price}
\end{align}
with a single shift $\delta$ that holds the forward~\eqref{eq:forward} exactly,
\begin{equation}
  \mu_i' = \mu_i - \delta, \qquad
  \delta = \log\!\Big(\textstyle\sum_j w_j e^{\mu_j+\frac12\sigma_j^2}\Big).
  \label{eq:delta}
\end{equation}
The Greeks follow from the same weighted sum. This makes the network a
\emph{forward density operator}: given $(\modelparams,T)$ it returns a terminal
density from which prices and sensitivities are analytic, and no-arbitrage
properties such as convexity in strike are inherited from the mixture rather than
imposed after the fact.

%% file: sections/04_mixture_surrogate.tex
\section{The Mixture Density Surrogate}
\label{sec:surrogate}

\begin{figure}[t]
  \centering
  \includegraphics[width=\textwidth]{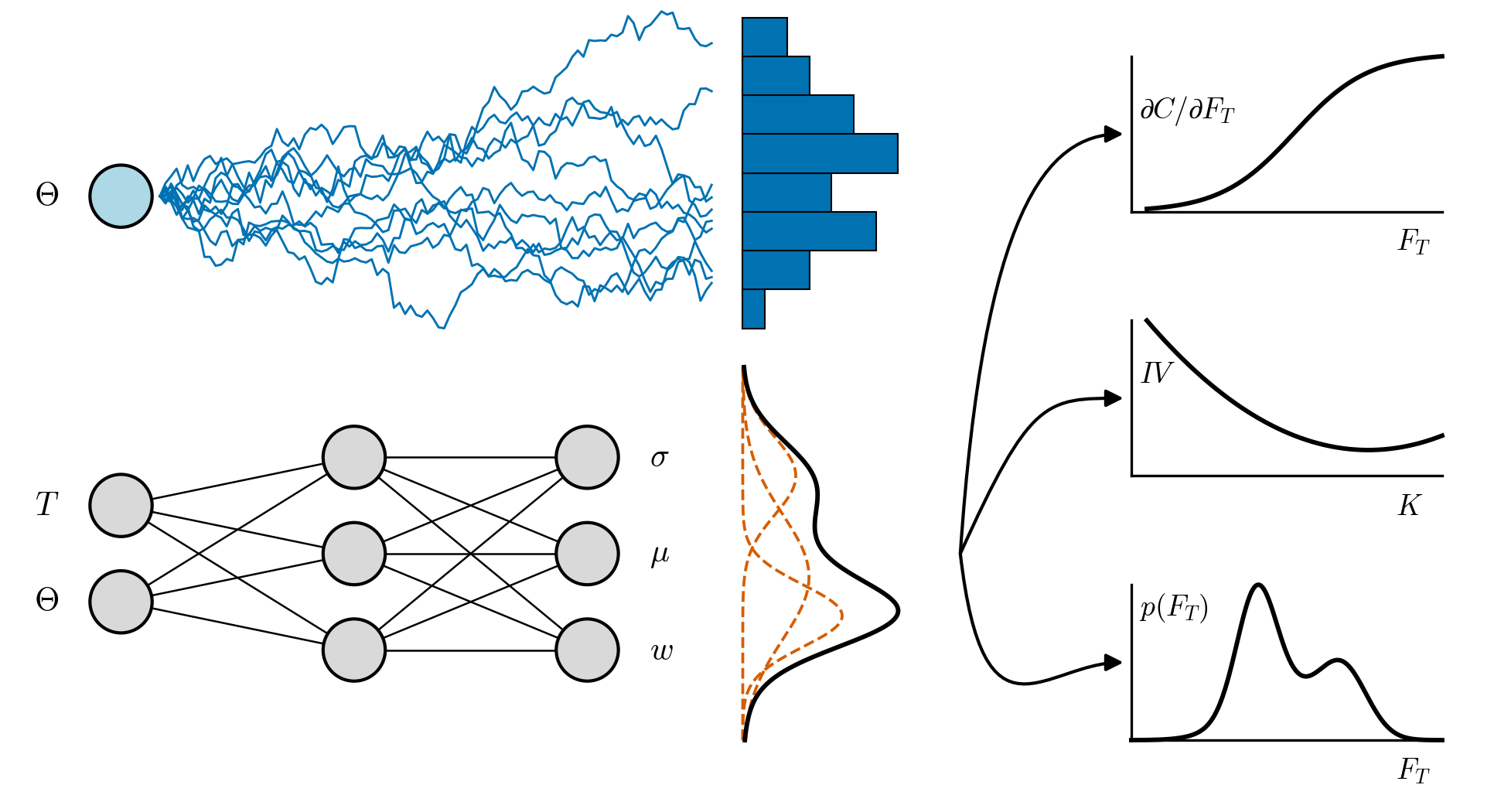}
  \caption{\textbf{Neural surrogate for return simulation.}
    \textit{Top:} Monte Carlo simulation generates return paths whose terminal
    values form an empirical density for each parameter set $\modelparams$.
    \textit{Bottom:} the MDN learns the direct mapping
    $(\modelparams,T)\mapsto\{w_i,\mu_i,\sigma_i\}$, producing the full terminal
    density in a single step, from which analytical option prices, implied
    volatilities and Greeks follow.}
  \label{fig:pipeline}
\end{figure}

The surrogate's task is to represent the terminal return density, replacing
simulation with a single forward pass: given the model parameters
$\modelparams$ and a maturity $T$, the network outputs a Gaussian mixture over
the (standardised) terminal return (Fig.~\ref{fig:pipeline}).

\paragraph{The mixture density network}

A distribution can be represented with a neural network in many ways (GANs,
diffusion models, and the like), but these would leave option pricing requiring
numerical integration of each payoff. We instead use a Mixture Density Network
(MDN) \citep{Bishop94}, which outputs the parameters of a Gaussian mixture. The
choice has three advantages: \emph{expressiveness} (a mixture approximates any
smooth density, improving systematically with more components); \emph{analytical
tractability} (each component prices in closed form via Black, so the option
value is a weighted sum, equation~\ref{eq:mixture_price}); and \emph{built-in
consistency} (a predicted density gives arbitrage-free prices automatically).
A handful of components already reprice listed S\&P~500 options within the
bid--ask spread; the trained surrogate uses a larger mixture
(Section~\ref{sec:errors}) to span the full range of GJR--GARCH shapes down to
the Monte Carlo accuracy floor. The MDN is a fully connected network mapping

\begin{equation}
  (w,\mu,\sigma) = f_{\mathrm{MDN}}(\modelparams, T \given \netparams),
  \label{eq:mdn_output}
\end{equation}

with learnable weights $\netparams$, yielding the density
$p_{\netparams}(X_T\given\modelparams,T)=\sum_{i=1}^M w_i\,\Normal(X_T\given\mu_i,\sigma_i^2)$.

\paragraph{Output heads}

For $M$ components the network emits three length-$M$ vectors (logits $\ell$,
means $\mu$, and raw scales $s$), converted to valid mixture parameters by

\begin{equation}
  w_i = \frac{e^{\ell_i}}{\sum_{j} e^{\ell_j}}, \qquad
  \mu_i \ \text{(linear head)}, \qquad
  \sigma_i = e^{s_i} + \sigma_{\min},
\end{equation}

with a small floor $\sigma_{\min}$ (e.g.\ $10^{-4}$) to avoid degenerate
components. In our implementation the heads use bounded transforms
($w=\mathrm{softmax}(6\tanh\cdot)$, $\sigma=\exp(5\tanh\cdot-3)$) for stable
training; the drift correction~\eqref{eq:delta} restores the forward at
pricing time.

%% file: sections/05_cdf_loss.tex
\section{Training Objective: CDF Matching}
\label{sec:loss}

\paragraph{Why likelihood underperforms}
The natural objective is to match the predicted density to the simulated one.
The standard way to do that is to minimise the negative log-likelihood (NLL) of
the simulated samples, which up to a constant equals the Kullback--Leibler
divergence from the simulated distribution to the model,
$\mathcal{L}_{\mathrm{NLL}}(\netparams) = -\tfrac1N\sum_j \log
p_{\netparams}(X_T^{(j)}\given\modelparams,T)$. This fits the density well but is
poorly aligned with pricing. A price is an integral of the payoff against the
density, not its height at any single point, so pricing depends on the
accumulated mass and a pointwise density fit is not the right target. The
mismatch is worse because the price weights returns by $e^x$: a small component
with large variance barely registers in the likelihood, yet that weighting lets
it move the price by a lot. A lognormal variant that fits the price density
directly,
$\mathcal{L}_{\mathrm{lognormal}}(\netparams) = -\tfrac1N\sum_j[\log
p_{\netparams}(X_T^{(j)}) - X_T^{(j)}]$, has the opposite flaw: likelihood
losses overweight rare tail events that are both under-sampled and nearly
irrelevant for pricing, injecting noise rather than signal.

\paragraph{Fitting the CDF, in quantile space}
From the Black formula~\eqref{eq:black}, once $F_0,K,r,\sigma$ are fixed the
option value is a functional of the return CDF. We therefore compare the
predicted and empirical CDFs at a fixed grid of quantile levels $q_1,\dots,q_Q$
uniformly spaced in $[0.001,0.999]$, rather than at fixed return values. Evaluating
 in quantile space gives equal weight to all probability mass and avoids the artificially small
errors of $x$-space evaluation in the tails (Fig.~\ref{fig:gmm_pdf_cdf}). For
each level $q$ the Monte Carlo threshold is $x_q=\cdfmc^{-1}(q)$, with empirical
CDF
\begin{equation}
  \mathrm{CDF}_{\mathrm{MC}}(x_q) = \frac1N\sum_{j=1}^N
  \mathbf{1}\{X_T^{(j)}\le x_q\},
  \label{eq:empirical_cdf}
\end{equation}
compared to the closed-form mixture CDF
$\mathrm{CDF}_{\mathrm{GMM}}(x_q)=\sum_{i=1}^M w_i\,\Phi(x_q;\mu_i,\sigma_i)$. The
loss is the root-mean-square deviation over the grid,
\begin{equation}
  \Lcdf(\netparams) = \sqrt{\frac1Q\sum_{q=1}^Q
  \big(\mathrm{CDF}_{\mathrm{GMM}}(x_q) - \mathrm{CDF}_{\mathrm{MC}}(x_q)\big)^2}.
  \label{eq:cdf_loss}
\end{equation}

\begin{figure}[t]
  \centering
  \includegraphics[width=\textwidth]{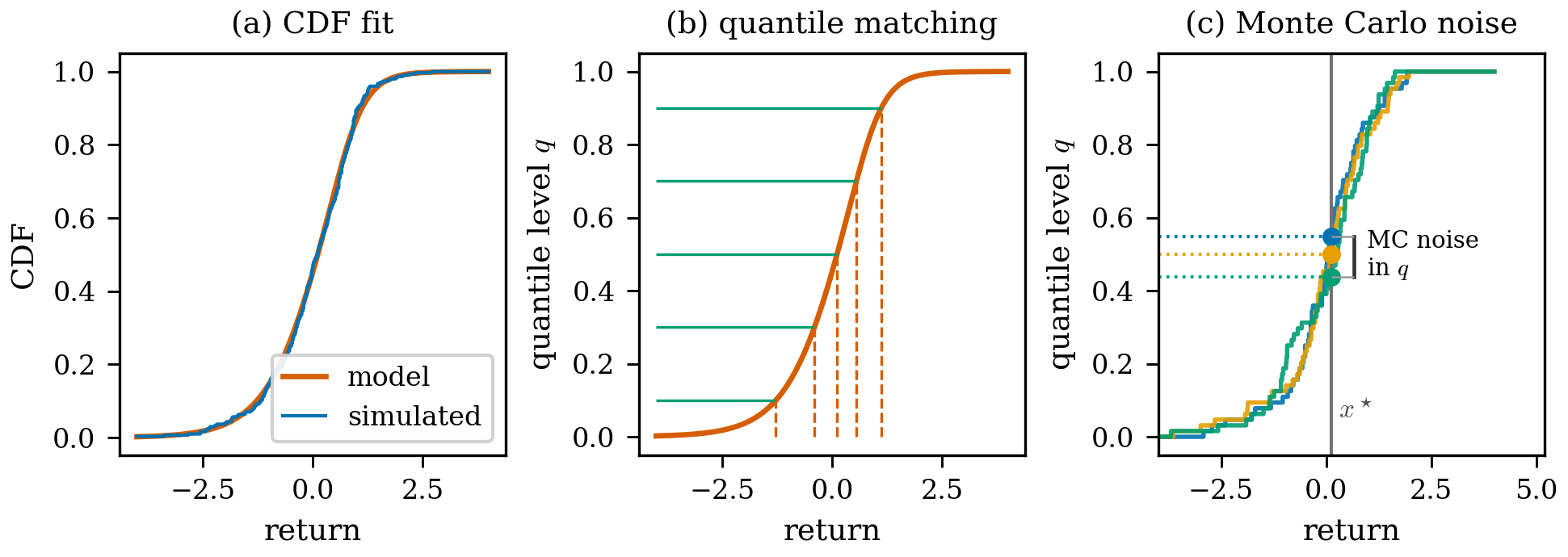}
  \caption{\textbf{Fitting the CDF.}
    \textit{(a)} Model CDF (red) versus the simulated CDF (blue).
    \textit{(b)} Quantile matching: target quantile levels (green) set Monte
    Carlo thresholds (red dashed) at which the two CDFs are compared.
    \textit{(c)} Monte Carlo noise: a few independent re-simulations of the same
    distribution give slightly different empirical CDFs, so reading them at one
    fixed return $x^\star$ returns different quantile levels $q$. That vertical
    spread is the accuracy limit analysed in Section~\ref{sec:floor}.}
  \label{fig:cdf_fitting}
\end{figure}

Conceptually related to Cram\'er-von Mises goodness-of-fit, this loss is smooth,
stable under gradient descent, and, as we now make precise, a direct proxy
for pricing error (Fig.~\ref{fig:cdf_fitting}). The targets are put on a common
scale across maturities by the dimensionless standardisation of
Section~\ref{sec:standardisation}, which leaves prices unchanged.

\paragraph{The CDF error bounds the pricing error}

Writing the call price~\eqref{eq:expectation_price} and integrating by parts in
the strike variable gives
\begin{equation}
  C(K) = e^{-rT}\!\int_{\ln(K/F_0)}^{\infty}\! F_0 e^{x}\big(1-G(x)\big)\,\dd x,
\end{equation}
where $G$ is the return CDF. The price thus depends on $G$ only through an
exponentially weighted integral of $1-G$. Hence for two return distributions with CDFs
$G_1,G_2$,

\begin{equation}
  \big|C_1(K)-C_2(K)\big| \le e^{-rT} F_0 \int e^{x}\,\big|G_1(x)-G_2(x)\big|\,\dd x
  \le \mathrm{const}\cdot \big\|G_1-G_2\big\|,
  \label{eq:price_bound}
\end{equation}

so a uniformly small CDF error implies a uniformly small pricing error across
strikes: minimising~\eqref{eq:cdf_loss} controls the quantity of interest, which is
why we train on the CDF rather than the density or the price. Section~\ref{sec:results}
confirms that the CDF loss comes within a small multiple of the noise floor at every quantile, far below the
likelihood objectives.

One refinement follows from the $e^{x}$ weight in the bound. That weight belongs to
the \emph{absolute} (dollar) price, which emphasises the upper tail. Matching the CDF
uniformly in quantile space instead targets the distribution itself, equally across all
probability mass. It is the model-agnostic primitive from which any payoff is priced.
Minimising the dollar error of a particular vanilla would call for a correspondingly
tail-weighted loss. We adopt the unweighted CDF as the default, since the surrogate's
purpose is a calibrated density rather than a single price.

%% file: sections/06_noise_floor.tex
\section{The CDF Sample-Noise Floor}
\label{sec:floor}
The training targets are themselves estimates: each empirical CDF
value~\eqref{eq:empirical_cdf} is computed from $N$ Monte Carlo samples and so
carries sampling noise. This puts a hard, model-independent floor on the test
error of \emph{any} surrogate: the targets pin down the truth only to within this
noise, so no surrogate can sit closer to the truth than the targets themselves
do. It is the benchmark against which we judge convergence.
\begin{proposition}[Distribution-free CDF noise floor]
\label{prop:floor}
Let $\cdfmc$ be the empirical CDF of $N$ i.i.d.\ draws and let quantile levels
$q$ be sampled uniformly on $(0,1)$ with thresholds $x_q=\cdfmc^{-1}(q)$. Then,
for any continuous return distribution, the per-quantile CDF estimate has standard deviation
\begin{equation}
  \operatorname{Std}\!\big[\mathrm{CDF}_{\mathrm{MC}}(x_q)\big]
  = \sqrt{\frac{q(1-q)}{N}},
  \label{eq:per_level_noise}
\end{equation}
and the expected root-mean-square CDF error over the quantile grid equals
\begin{equation}
  \Lmin = \sqrt{\E_q\!\left[\frac{q(1-q)}{N}\right]} = \sqrt{\frac{1}{6N}},
  \qquad\text{since}\quad \int_0^1 q(1-q)\,\dd q = \tfrac16.
  \label{eq:floor}
\end{equation}
\end{proposition}
Equation~\eqref{eq:per_level_noise} follows because $\mathrm{CDF}_{\mathrm{MC}}(x_q)$
is the mean of $N$ Bernoulli$(q)$ indicators; equation~\eqref{eq:floor} is the
uniform-$q$ average. The result is \emph{distribution-free}: it holds regardless
of tails or skewness. At the $N=10^7$ paths used for our targets, this floor is
$\Lmin\approx1.29\times10^{-4}$, the value the rest of the paper measures against.
\paragraph{Uncertainty is concentrated at the money}
The per-quantile noise~\eqref{eq:per_level_noise} is largest at $q=\tfrac12$ (at the
money) and vanishes as $q\to0,1$, so Monte Carlo pricing uncertainty is
concentrated near the money rather than in the tails. The floor is in fact lowest
in the tails: at the $99.9\%$ quantile it is more than an order of magnitude below
its grid average, and the surrogate's error there stays small in absolute terms,
so value-at-risk and other tail-risk measures remain well within reach. The one
visible departure, the extreme-quantile spike of Section~\ref{sec:results_loss},
is a capacity bias rather than noise: a finite Gaussian mixture has lighter tails
than the true density, leaving a small, fixed residual that the collapsing floor
magnifies in floor units while it stays negligible in absolute terms and for
pricing.
\paragraph{Worst-case error and its growth with usage}
A single case is summarised by its 99th-percentile error, the level exceeded once
in a hundred, which for mean-zero Gaussian noise is $2.58$ times the per-quantile
standard deviation. This worst case grows with usage: a calibration or risk run
meets the largest error among many. The expected maximum over $M$ independent
cases exceeds the per-case standard deviation $\sigma$ by an extreme-value factor,
\begin{equation}
  \text{max error} \approx \sigma\sqrt{2\ln M}, \qquad
  \sigma \approx \max_q \sqrt{\frac{q(1-q)}{N}} = \frac{1}{2\sqrt{N}},
  \label{eq:max_error}
\end{equation}
which grows only as $\sqrt{\ln M}$: even across a million priced surfaces the
expected worst case is about $5\,\sigma$, so heavy use keeps it within a small
multiple of the floor.
\paragraph{Centering tightens the floor}
Centering the targets lowers the floor further, giving two reference levels rather
than one. Subtracting the sample mean before forming the CDF, as we do when
preparing the targets, reduces the large-$N$ variance at the quantile thresholds.
For Gaussian targets the per-quantile variance at $x_q=\Phi^{-1}(q)$ drops from
$q(1-q)/N$ to approximately $\big(q(1-q)-\phi(x_q)^2\big)/N$, with $\phi$ the
standard normal density; unlike the distribution-free floor above, this gain
depends on the target distribution. The reduction $\phi(x_q)^2/N$ is largest near
the median and fades in the tails, so centering flattens the error curve in the
middle, improving both the average and, through~\eqref{eq:max_error}, the
worst-case error exactly where pricing is most sensitive. The two floors are then
the distribution-free pure-MC floor $\Lmin$, which any set of $N$-path targets
must respect, and the tighter centered floor that mean-centered targets actually
permit. A surrogate whose validation loss approaches either has saturated the
available training signal: no change of architecture can do better without more
simulation paths.

%% file: sections/07_error_anatomy.tex
\section{The Four Sources of Error}
\label{sec:anatomy}

The out-of-sample error splits into four sources, each with a control that sets
it and an observable that reveals it. Writing the test error as a sum of
approximately independent parts,
\begin{equation}
  \E\!\big[\varepsilon_{\text{test}}^2\big]\;\approx\;
  \underbrace{\Lmin^2}_{\text{label noise}}\;+\;
  \underbrace{e_{\mathrm{cap}}^2}_{\text{capacity}}\;+\;
  \underbrace{e_{\mathrm{cov}}^2}_{\text{coverage}}\;+\;
  \underbrace{e_{\mathrm{opt}}^2}_{\text{optimiser}},
  \label{eq:anatomy}
\end{equation}
each part has its own cause and signature (Table~\ref{tab:anatomy}). This is the
organising idea of the paper: a surrogate with \emph{known} error is one in which
every term has been pushed below the target precision and \emph{verified} to be
there.

\begin{table}[t]
\centering
\small
\caption{The four contributions to the out-of-sample error, the control that sets each,
and the signal that reveals it. The whole modelling task is to push all four
below the target precision $\varepsilon$ and confirm it.}
\label{tab:anatomy}
\begin{tabularx}{\linewidth}{@{}l Y Y@{}}
\toprule
\textbf{Error term} & \textbf{Set by} & \textbf{How you see it} \\
\midrule
Label noise $\Lmin=\sqrt{1/(6N)}$ &
  Monte Carlo paths per case $N$ &
  the floor itself; it fixes the target \\[2pt]
Capacity bias $e_{\mathrm{cap}}$ &
  network architecture (expressiveness) &
  test error stays above the floor however much data is added \\[2pt]
Coverage variance $e_{\mathrm{cov}}$ &
  number and placement of cases (Sobol) &
  data-efficiency curve; the train/test gap \\[2pt]
Optimiser temperature $e_{\mathrm{opt}}\!\approx\!\sqrt{c_{\mathrm{opt}}\,\eta/B}$ &
  learning rate and batch, through $\eta/B$ &
  late-phase jitter; vanishes as $\eta\to0$ \\
\bottomrule
\end{tabularx}
\end{table}

\paragraph{The four terms}
The \emph{label noise} $\Lmin=\sqrt{1/(6N)}$ is the irreducible Monte Carlo error
in the targets (Section~\ref{sec:floor}); it is the finest precision the data
could ever support. The \emph{capacity bias} $e_{\mathrm{cap}}$ is the error of
the best fit the network can represent. A sufficiently expressive network can
approximate the smooth price map arbitrarily well in principle, by universal
approximation \citep{Hornik1991}, but one that is too small or too smooth for the
true map leaves this term positive no matter how much data we provide, and only a
better architecture removes it (Section~\ref{sec:errors}). The \emph{coverage
variance} $e_{\mathrm{cov}}$ comes from sampling the parameter space too sparsely
or unevenly; the surrogate predicts by interpolating nearby cases, so this term
shrinks as cases are added, and faster with low-discrepancy sampling
(Section~\ref{sec:data}). The \emph{optimiser temperature}
$e_{\mathrm{opt}}\approx\sqrt{c_{\mathrm{opt}}\,\eta/B}$ is the residual jitter of
stochastic training as it orbits the best fit; it depends only on the per-sample
noise scale $\eta/B$ and is the one term the learning-rate schedule can remove
(Sections~\ref{sec:errors} and~\ref{sec:design}).

\paragraph{Reading the terms without the truth}
We never observe the true price map, yet each term leaves a fingerprint we can
measure, and Section~\ref{sec:results} reads them one at a time. Three leave a
direct fingerprint: the floor from the path count
(Section~\ref{sec:results_accuracy}); the coverage variance from the
data-efficiency curve and the closing train/test gap
(Section~\ref{sec:results_data}); and the optimiser temperature from the
amplitude of the late-phase jitter and the schedule comparison
(Section~\ref{sec:errors} and Fig.~\ref{fig:schedule}). The capacity bias is then
whatever the test error retains once added data and training stop lowering it,
pinned by subtraction (Section~\ref{sec:results_frontier}). Confirming that these
contributions account for the observed test error validates the decomposition.

%% file: sections/08_design_rules.tex
\section{Design Rules}
\label{sec:design}

The noise floor of Section~\ref{sec:floor} does more than bound the error; it also
tells us how to set training up to reach it. Suppose we want a target precision
$\varepsilon$ on the out-of-sample CDF error. Three of the four error terms of
Section~\ref{sec:anatomy} are set by training choices: the paths per case (label
noise), the number of parameter cases (coverage), and the learning-rate schedule
(optimiser temperature). The fourth, capacity, is an architecture choice taken up
later (Section~\ref{sec:errors}). The floor fixes each of the three.

\paragraph{Paths per case fix the target}
The test labels are themselves Monte Carlo estimates from $N$ paths, so the
smallest error any model can reach is the floor $\Lmin=\sqrt{1/(6N)}$. To aim for
$\varepsilon$ we therefore need
\[
  N \;\ge\; \frac{1}{6\,\varepsilon^{2}}.
\]
For example, $\varepsilon=2\times10^{-4}$ needs about $4\times10^{6}$ paths per
case, and $\varepsilon=10^{-4}$ about $1.7\times10^{7}$. This cost is
unavoidable: only more paths move the target lower.

\paragraph{Cases fix the coverage}
The number of parameter cases $N_{\mathrm{cases}}$ sets how densely we sample the
parameter space. The network predicts at a new $\Theta$ by interpolating between
nearby training cases, so its test error has two parts,
\[
  \varepsilon_{\text{test}}^{2}\;\approx\;\Lmin^{2}\;+\;\varepsilon_{\text{cover}}^{2}(N_{\mathrm{cases}}),
\]
the floor plus an extra term that grows when the space is sampled too sparsely.
How fast $\varepsilon_{\text{cover}}$ shrinks as cases are added depends on how
smoothly prices vary with $\Theta$, which is not known ahead of time. A
low-discrepancy (Sobol) design spreads cases without gaps and cannot hurt, a
useful default as the parameter dimension grows. The stopping rule is empirical
(Section~\ref{sec:data}): add cases until the test error stops improving and the
training and test errors agree.

\paragraph{The final learning rate fixes the residual wobble}
Stochastic training settles into a small, random orbit around the best fit. The
orbit grows with the learning rate $\eta$ and shrinks with the batch size $B$:
the minibatch gradient is an average of $B$ independent per-sample gradients, so
its variance falls as $1/B$. The orbit therefore adds an optimiser error
\[
  e_{\mathrm{opt}}(\eta)\;\approx\;\sqrt{c_{\mathrm{opt}}\,\frac{\eta}{B}},
\]
where $c_{\mathrm{opt}}$, a constant we measure (Section~\ref{sec:errors}),
absorbs the curvature and Adam's preconditioning. The best final rate
$\eta^\star$ is the one whose orbit is just small enough to sink beneath the
floor,
\[
  e_{\mathrm{opt}}(\eta^\star)\approx \Lmin
  \quad\Longrightarrow\quad
  \eta^\star \approx \frac{B}{c_{\mathrm{opt}}}\,\Lmin^{2},
\]
which we find to be about $6\times10^{-5}$ (Section~\ref{sec:errors}). A higher
final rate leaves a visible wobble above the target; a lower one shrinks the
wobble below a noise we cannot even measure, and only adds training time.

\paragraph{How long to train, and why to start fast}
The loss settles at a rate set by the learning rate: reaching the orbit at rate
$\eta$ takes about $a/\eta$ samples, with $a$ measured in
Section~\ref{sec:errors}. Because $\eta^\star$ is small, this final phase is the
main cost, taking a few times $a/\eta^\star$ samples. The early part of training is
much cheaper: a high learning rate clears most of the loss in only about $1/\eta$
steps. This is why we start high, for fast initial progress, and anneal down to
$\eta^\star$ for the precise endgame. Plotted against samples, such a schedule
follows the best result reachable at any moment, namely the lower edge of the
family of constant-rate curves (Section~\ref{sec:errors}). Training is done when this
curve flattens out.

\paragraph{The recipe}
To reach precision $\varepsilon$:
\begin{enumerate}[leftmargin=*]
  \item simulate $N \ge 1/(6\varepsilon^{2})$ paths per case;
  \item add Sobol cases until the test error stops improving and matches the
        training error;
  \item anneal the learning rate from a high starting value down to
        $\eta^\star\approx (B/c_{\mathrm{opt}})\,\Lmin^{2}$, training until the test
        loss flattens out.
\end{enumerate}
Once the out-of-sample CDF error sits at $\sqrt{1/(6N)}$, the surrogate is as accurate
as the training data allow. More steps, larger networks, or smaller learning
rates then change nothing; only more Monte Carlo paths can do better, and the
modelling task is complete.

%% file: sections/09_gjr_garch_model.tex
\section{The GJR--GARCH Model and Its Reduced Form}
\label{sec:model}

The GJR--GARCH(1,1) model \citep{glosten1993relation} extends the classical
GARCH specification with an asymmetric term that amplifies volatility after
negative shocks, capturing the empirical \emph{leverage effect}. It is defined by
\begin{equation}
  \begin{aligned}
    x_t &= \mu + \epsilon_t, \qquad
    \epsilon_t = \sigma_t z_t, \qquad
    z_t \sim \skewt_{\nu,\lambda}(0,1),\\[3pt]
    \sigma_t^2 &= \omega + \alpha\,\epsilon_{t-1}^2
      + \gamma\,\epsilon_{t-1}^2\,\mathbf{1}\{\epsilon_{t-1}<0\}
      + \beta\,\sigma_{t-1}^2,
  \end{aligned}
  \label{eq:gjr}
\end{equation}
where $x_t$ is the one-period log-return, $\mu$ the mean return, $\omega>0$ the
variance intercept, $\alpha\ge 0$ the reaction to squared shocks, $\gamma\ge 0$
the asymmetry (leverage) coefficient, $\beta\ge 0$ the persistence term, and
$\sigma_0^2>0$ the initial variance. The innovations $z_t$ follow Hansen's
skewed Student-$t$ distribution $\skewt_{\nu,\lambda}(0,1)$
\citep{hansen1994autoregressive}, standardised to zero mean and unit variance,
with $\nu>2$ controlling tail thickness and $\lambda\in(-1,1)$ controlling
skewness (Appendix~\ref{app:gjr}).

\paragraph{Long-run variance and persistence}
Define the expected contribution of negative shocks to total variance,
\begin{equation}
  p_- := \E\!\left[z_t^2\,\mathbf{1}\{z_t<0\}\right],
\end{equation}
which depends on the innovation parameters $(\nu,\lambda)$ and equals
$p_-=\tfrac12$ for symmetric innovations ($\lambda=0$). Taking expectations of the
variance recursion~\eqref{eq:gjr} gives the linear relation
$\E[\sigma_t^2]=\omega+\kappa\,\E[\sigma_{t-1}^2]$, with persistence coefficient
\begin{equation}
  \kappa = \alpha + \beta + \gamma\,p_-, \qquad \kappa < 1,
  \label{eq:persistence}
\end{equation}
and $\kappa<1$ the covariance-stationarity condition. Solving the recursion gives
the variance conditioned on the initial level $\sigma_0^2$,
\begin{equation}
  \E[\sigma_t^2] = v + \kappa^{t}\big(\sigma_0^2 - v\big), \qquad
  v := \frac{\omega}{1-\kappa},
  \label{eq:condvar_phys}
\end{equation}
exact for every $t$; it relaxes at rate $\kappa$ to the unconditional (long-run)
variance $v$, the stationary value reached as $t\to\infty$.

\paragraph{Shift and scale symmetries}
The model has two elementary invariances: (i) adding a constant to all
returns shifts $\mu$, and (ii) multiplying all returns by a scalar rescales
$\omega$, $\sigma_0^2$, and $\sigma_t$ but leaves the dynamics unchanged.
Different parameterisations can thus describe statistically equivalent processes.
A surrogate trained on raw parameters could assign inconsistent valuations to
such equivalent configurations, introducing artificial arbitrage. We eliminate
this redundancy with a dimensionless coordinate system.

\begin{figure}[t]
  \centering
  \includegraphics[width=\textwidth]{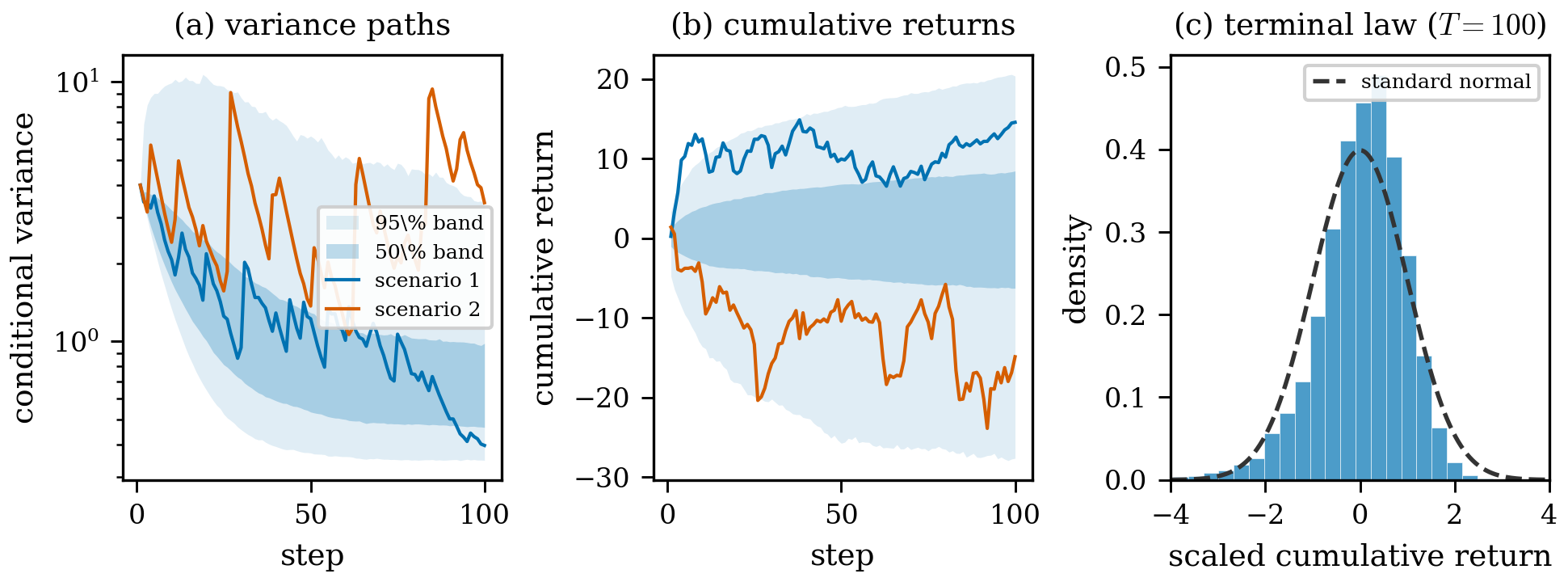}
  \caption{Monte Carlo illustration of the dimensionless reduced-form
    GJR--GARCH model. \textit{Left:} variance paths from an elevated initial
    level, on a logarithmic scale with 50\% and 95\% confidence intervals.
    \textit{Middle:} cumulative return trajectories. \textit{Right:} the scaled
    terminal return distribution at $T=100$; the dashed curve is a standard normal
    for comparison, showing the heavier tails and slight skew of the GJR--GARCH
    distribution.}
  \label{fig:garch_reduced_demo}
\end{figure}

\paragraph{Dimensionless reduced form (DRF)}
The DRF normalises both the unconditional mean and variance,
\begin{equation}
  x_t' = \frac{x_t-\mu}{\sqrt{v}}, \qquad
  (\sigma_0')^2 = \frac{\sigma_0^2}{v}, \qquad
  \omega' = 1-\kappa,
\end{equation}
so that the reduced process has zero mean and unit long-run variance. The
reduced parameter vector
\begin{equation}
  \theta_{\DRF} = \big(\alpha,\gamma,\beta,(\sigma_0')^2,\nu,\lambda\big)
  \label{eq:theta}
\end{equation}
is the concrete form of the network input $\Theta$ from Part~I, and fully
determines the dimensionless \emph{shape} of the return distribution. The DRF
enforces the model's invariances analytically, so the network learns only
genuine shape dependencies, with three benefits: \emph{(i) arbitrage
consistency}: processes that differ only by a shift or rescaling map to the same
reduced coordinates, so they are priced identically by construction;
\emph{(ii) dimensional reduction}: removing the redundant mean and variance
scales leaves a smaller set of shape variables, reducing the data needed for full
coverage; and \emph{(iii) numerical robustness}: dimensionless quantities of
order unity improve the conditioning of both simulation and optimisation.
Figure~\ref{fig:garch_reduced_demo} illustrates the reduced-form dynamics.

\paragraph{Back mapping}
To recover physical-scale parameters for a target mean $\mu^\star$ and variance
$v^\star>0$,
\begin{equation}
  x_t = \mu^\star + x_t'\sqrt{v^\star}, \qquad
  \omega = (1-\kappa)\,v^\star, \qquad
  \sigma_0^2 = (\sigma_0')^2\,v^\star,
\end{equation}
while $(\alpha,\beta,\gamma,\nu,\lambda)$ remain unchanged. All simulation and
neural density models operate in DRF coordinates; reconstruction to physical
units is applied only when pricing or interpreting results.

\paragraph{Cumulative-variance standardisation}
\label{sec:standardisation}
Rather than modelling each one-period return as in simulation, we predict the
standardised cumulative return over the horizon. The reduced form fixes the
\emph{long-run} variance to one, but the initial variance $(\sigma_0')^2$ is a
free coordinate of $\theta_{\DRF}$ and need not equal it. In reduced coordinates
the conditional variance~\eqref{eq:condvar_phys} becomes
\begin{equation}
  \E[(\sigma_t')^2] = 1 + \kappa^{t}\big((\sigma_0')^2 - 1\big)
  \label{eq:condvar}
\end{equation}
The return increments are serially uncorrelated, so their variances add, and the
cumulative variance of the $T$-step return is
\begin{equation}
  s_T^2 \;:=\; \sum_{t=0}^{T-1}\E[(\sigma_t')^2]
        \;=\; T + \big((\sigma_0')^2-1\big)\,\frac{1-\kappa^{T}}{1-\kappa}.
  \label{eq:cumvar}
\end{equation}

We standardise by this exact cumulative standard deviation,
\begin{equation}
  X_T = \frac{1}{s_T}\sum_{t=1}^{T} x_t',
  \label{eq:target}
\end{equation}
so that $X_T$ has unit variance at \emph{every} horizon, placing all maturities
on a genuinely common scale for a single network to generalise across. When the
initial variance sits at the long-run level ($(\sigma_0')^2=1$), or as
$T\to\infty$, the bracket in~\eqref{eq:cumvar} vanishes, $s_T^2\to T$, and
\eqref{eq:target} reduces to the naive $1/\sqrt{T}$ scaling. The correction
matters when the initial variance is away from its long-run level
($(\sigma_0')^2\ne1$) and the horizon is short relative to the mixing time
$1/(1-\kappa)$, i.e.\ in the high-persistence regime. There the naive
$1/\sqrt{T}$ scaling would leave the standardised variance off by as much as a
factor $(\sigma_0')^2$, at the shortest horizon.

%% file: sections/10_training_data.tex
\section{Training Data and Parameter Sampling}
\label{sec:data}

This section describes the training set: how many parameter cases the surrogate
sees, and where they sit in the parameter space. These two choices are what set
the coverage term $e_{\mathrm{cov}}$ of the decomposition~(\ref{eq:anatomy}). We
place the cases with a low-discrepancy (Sobol) design, which fills the space
without the clustering and gaps of plain random sampling (Appendix~\ref{app:data})
and matters more as the dimension grows.

The dataset is generated by Monte Carlo simulation of the reduced-form
GJR--GARCH model and published on the Hugging Face Hub as
\texttt{simu-ai/garch\_densities}; the experiments in this paper consume it
directly. Each entry pairs a parameter set $\theta_{\DRF}$ and a maturity $T$
with the empirical quantiles $\{x_q\}$ of the standardised terminal return $X_T$.
Storing quantiles rather than raw paths gives a compact representation that is
directly relevant for pricing across strikes (Section~\ref{sec:loss}); each
entry is thus a tuple $(\theta_{\DRF}, T, \{x_q\})$ on the same $Q=512$ quantile
grid the CDF loss is evaluated on (Section~\ref{sec:loss}), uniformly spaced in
$[0.001,0.999]$.

\paragraph{Selecting parameter cases}
Parameters are drawn from a Sobol sequence \citep{Sobol1967} over the admissible
region (illustrated in Appendix~\ref{app:data}). The six free parameters of
$\theta_{\DRF}$ are sampled as:
\begin{itemize}[leftmargin=*,itemsep=0.25em]
  \item \textbf{Degrees of freedom $\nu$:} logarithmic in $[4,100]$.
  \item \textbf{Skew $\lambda$:} uniform in $[-0.98,0.98]$.
  \item \textbf{Initial variance $(\sigma_0')^2$:} logarithmic in $[1/30,30]$.
  \item \textbf{Persistence $(\alpha,\gamma\,p_-,\beta)$:} uniform on the simplex
    with $\alpha+\gamma\,p_-+\beta<0.95$ (so $\kappa<0.95$,
    equation~\ref{eq:persistence}). Three sorted uniforms
    $u_{(1)}\le u_{(2)}\le u_{(3)}$ give $\alpha=u_{(1)}$,
    $\gamma\,p_-=u_{(2)}-u_{(1)}$, $\beta=u_{(3)}-u_{(2)}$, after which $\gamma$
    is recovered from the $(\nu,\lambda)$-dependent $p_-$.
\end{itemize}
We sample the persistence quantity $\gamma\,p_-$ rather than $\gamma$ directly,
consistently with the stationarity coefficient
$\kappa=\alpha+\beta+\gamma\,p_-$, so the simplex is laid out in the same
coordinates that govern stationarity; no separate bound on $\gamma$ is needed,
since stationarity is controlled through the product $\gamma\,p_-$.

\paragraph{Train/test split and scale}
We use the dataset's published train/test split as provided. Both splits are
\emph{independent} low-discrepancy (Sobol) designs over the parameter space,
generated from non-overlapping streams so that no test configuration coincides
with a training one; the test set is therefore strictly out-of-sample, and every
error we report is measured on it. The training set covers $2^{17}\approx131\text{k}$ parameter
cases and the test set $2^{15}\approx32\text{k}$ cases, each simulated with
$N=10^{7}$ paths over a $1{,}000$-step horizon. Every case is paired with a
sample of maturities spanning that horizon ($\approx50$--$120$ per case, the
shortest horizons densely and longer ones sampled), so the dataset contains
about $7.5$M training and $1.9$M test density curves in total. Generation ran on
rented GPU servers over several days.

\paragraph{Accelerating the skewed-$t$ sampler}
The slowest step in simulation is drawing from the Hansen skewed-$t$, which
requires many inverse-CDF evaluations. We precompute, once per $(\nu,\lambda)$, a
dense inverse-CDF lookup table: a uniform grid $u\in(10^{-6},1-10^{-6})$ is
transformed to skewed-$t$ values in a single vectorised pass and the pairs
$(u_k,x_k)$ are stored on device. At runtime, sampling draws $u\sim U(0,1)$ and
linearly interpolates in the table. Precision is controlled by the grid size
($\sim10^{5}$ points), with endpoints clipped to keep the table monotone.

%% file: sections/11_capacity_optimisation.tex
\section{Capacity and Optimisation}
\label{sec:errors}
This section covers the two training-side choices: the network architecture and
the learning-rate schedule. They set the last two error terms of the
decomposition~(\ref{eq:anatomy}), the capacity bias $e_{\mathrm{cap}}$ and the
optimiser temperature $e_{\mathrm{opt}}$. We also measure here the constants that
the Section~\ref{sec:design} design rules rely on. Checking that all four error
terms add up to the test error is left to Section~\ref{sec:results}.

\paragraph{Architecture and optimisation}
The surrogate is a small fully connected MDN: one to four hidden layers of $32$
to $512$ units (plus one $768$-wide probe), arranged as uniform trunks or tapered
funnels, feeding three output heads that produce a mixture of $16$ to $128$
Gaussian components. We train it with Adam on the CDF loss~\eqref{eq:cdf_loss}
over the $Q=512$ quantile grid, using minibatches of a few hundred cases and the
hold-then-decay schedule below, for $5\times10^{7}$ to $10^{8}$ presented cases.

The inputs are the reduced-form parameters of equation~\eqref{eq:theta} plus the
maturity $T$. We log-transform $\nu$, $(\sigma_0')^2$ and $T$, and add two
features aimed at the hardest regions: $\log(1.01-\kappa)$, which stretches out
the near-nonstationary corner where the density changes fastest, and flags for
the shortest maturities $T\in\{1,2,3\}$. Full settings are in
Appendix~\ref{app:opt}.

\paragraph{Activations and topology}
Of the architecture choices, mixture size matters most: adding components lowers
the test error, steeply up to about sixty-four. Network shape comes next: a trunk
that narrows toward the output (a tapered funnel) beats a uniform-width trunk of
the same size. Smooth activations (GELU) worked best in earlier screening.
Section~\ref{sec:results} gives the full ranking and the trade-off between
accuracy and speed.

\paragraph{Learning-rate dynamics}
A high learning rate learns fast at first, but the noisy gradients keep it
bouncing around a good solution instead of settling into one. A low rate is
slower to start but settles lower. A decaying schedule gets both: a fast start,
then gradual settling (Fig.~\ref{fig:schedule}).

At the full $10^{8}$-sample budget the constant-high rate stalls at $4.5\times$
the floor, while all three decaying schedules land in a tight band just above it:
cosine at $1.11\times$, geometric at $1.13\times$, and hold-then-$1/t$ at
$1.16\times$. What matters is decaying at all, far more than the exact shape.
These runs all use a uniform $[512,512]$ trunk, so even the best of them sits a
little above the headline $[512,256,128]$ taper's $1.07\times$
(Section~\ref{sec:results}); the schedule sweep predates the taper and was not
re-run on it.

In practice, hold a high rate to capture the structure, then decay to a small
final rate, just low enough for the wobble to drop below the floor and no lower
(the design rule of Section~\ref{sec:design}).

\paragraph{Convergence to the noise floor}
Trained to convergence, the test error drops toward the Monte Carlo floor of
Section~\ref{sec:floor} and flattens out at a small multiple of it
(Fig.~\ref{fig:schedule}a). On the \emph{training} targets it can even dip below
the floor. That is possible because prices vary smoothly with the parameters, so
the network fits a smooth surface through the noisy per-case targets, and that
surface is more accurate than the noisy points it passes through. On the
\emph{independent} test set the error settles just above the floor, never below
it. That is what you see when a model has learned the true distribution rather
than memorised the sampling noise.

%% file: sections/12_results.tex
\section{Accuracy and Error Analysis}
\label{sec:results}

This section reports what the trained surrogate achieves, in three parts: first,
its test accuracy, and whether the leftover error is just sampling noise or a real
fault in the model; then, what controls that error, checking the four-term
decomposition of Section~\ref{sec:anatomy} term by term; and finally, how fast it
prices against simulation. The behaviour of the priced surfaces is taken up in
Section~\ref{sec:behaviour}.

\subsection{Accuracy and the Noise Floor}
\label{sec:results_accuracy}
The out-of-sample CDF error comes within $10\%$ of the Monte Carlo floor of
Section~\ref{sec:floor}. The best configuration reaches a mean CDF error of
$1.38\times10^{-4}$, against the $N=10^{7}$ floor of $1.29\times10^{-4}$, about
$1.07\times$ the floor. That configuration is a tapered $[512,256,128]$ trunk with
a $128$-component mixture and the reduced-form features of
Section~\ref{sec:errors}, trained under the cosine schedule of
Section~\ref{sec:design} at the full $10^{8}$-sample budget.

\begin{figure}[t]
  \centering
  \includegraphics[width=\textwidth]{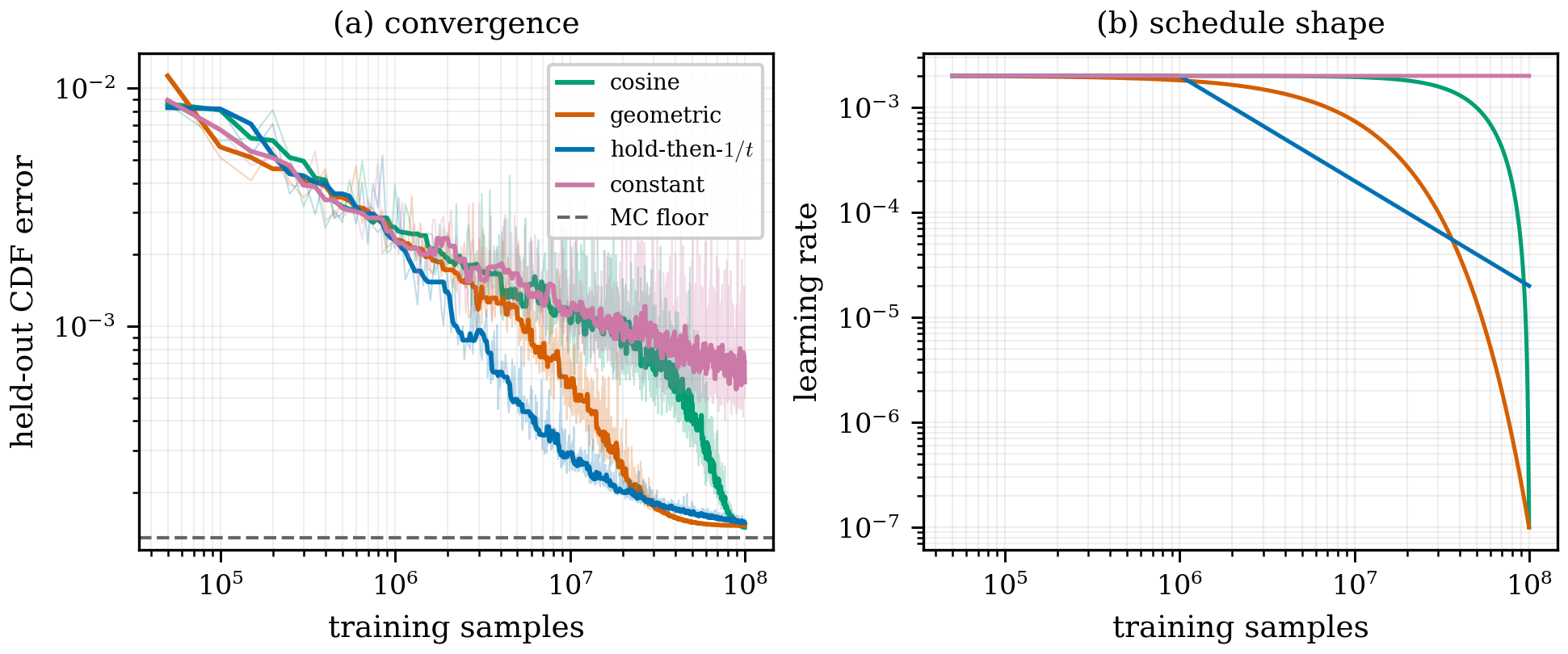}
  \caption{Learning-rate schedule and convergence. \textit{(a)} Test CDF error
    versus presented samples (faint: raw; bold: rolling median), with the Monte
    Carlo floor dashed. \textit{(b)} The schedule that produced each curve. A
    constant-high rate remains noisy and well above the floor; the decaying schedules
    converge far lower and into the same neighbourhood, so the choice among them is
    second-order next to decaying at all.}
  \label{fig:schedule}
\end{figure}

As training proceeds the error falls steadily and flattens onto this small
multiple of the floor (Fig.~\ref{fig:schedule}a), with the longest budget getting
closest. This is what you see when the limit is set by label noise, not by fitting
or coverage. Measured against the tighter centered floor that mean-centered
targets permit (Section~\ref{sec:floor}), the same error is about $1.45\times$.
Either way it is well within tolerance for most uses.

\subsection{Residual Error across the Parameter Space}
\label{sec:results_map}
\begin{figure}[t]
  \centering
  \includegraphics[width=\textwidth]{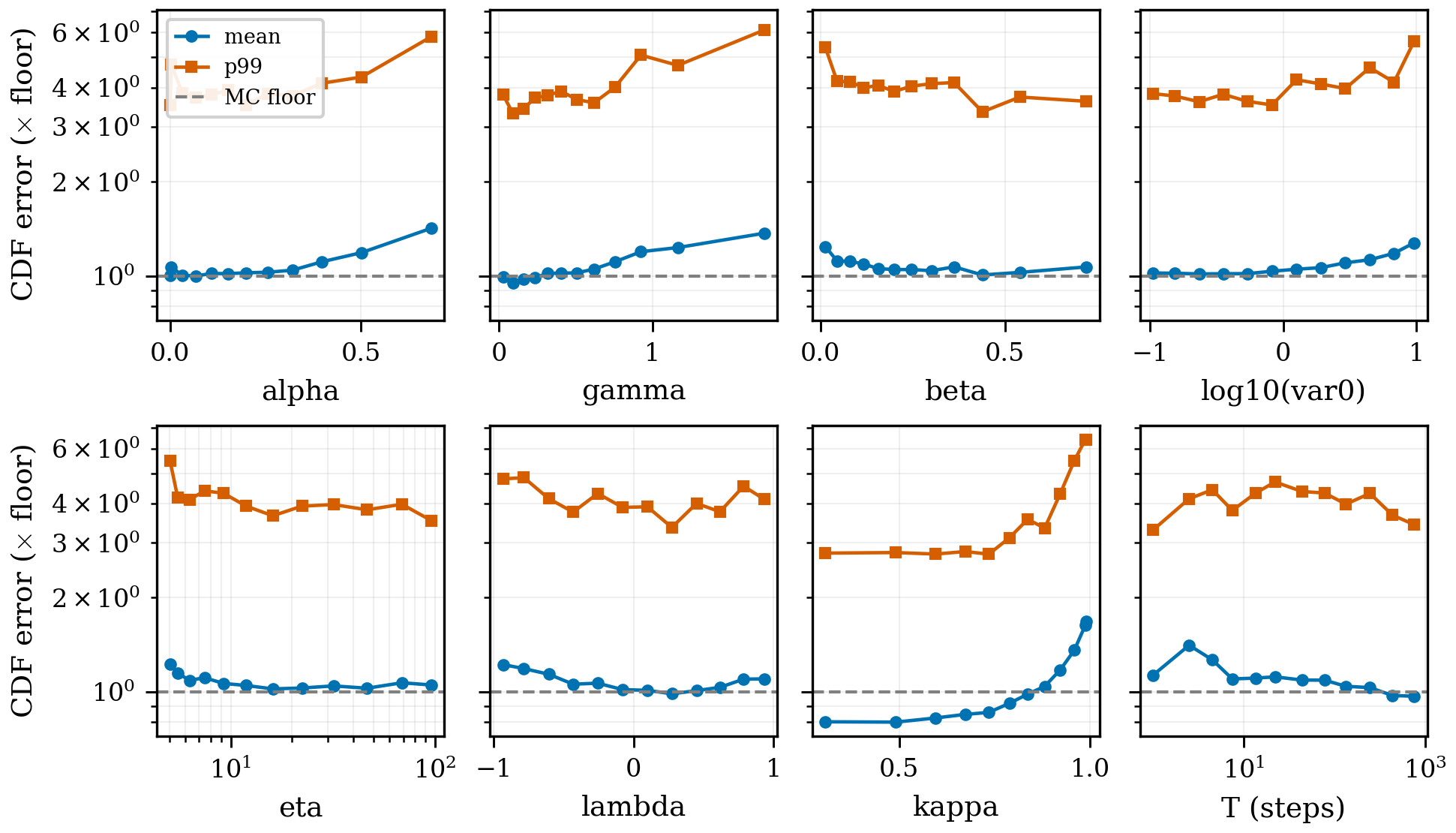}
  \caption{Test CDF error across the parameter space for the best model: per-bin
    mean and $99$th percentile ($\times$ floor) over equal-count bins of each
    parameter and maturity. The error is largest in the high-persistence corner and
    at the shortest maturities, where the per-bin mean stays within about
    $1.7\times$ the floor and the $99$th percentile within about $6\times$.}
  \label{fig:error_map}
\end{figure}

The residual above the floor is not spread evenly across the parameter space.
Figure~\ref{fig:error_map} bins the test error across the GJR--GARCH parameters
and maturity and reports the per-bin mean and $99$th percentile. Two regions carry
most of the error: the near-nonstationary corner, where persistence $\kappa\to1$
and the terminal density changes fastest, and the shortest maturities
$T\in\{1,2,3\}$, where the terminal return is the sum of only a few heavy-tailed
steps.

Even these hard regions stay close to the floor. Thanks to the reduced-form
standardisation, the persistence and short-horizon features, and enough capacity,
the mean error in the worst bins (the $\kappa\!\to\!1$ corner) stays within about
$1.7\times$ the floor, against $1.0$--$1.2\times$ across the easy interior; even
the $99$th-percentile case there reaches only about $4$--$6\times$ the floor. A
closer look shows the error sits in a small fraction of cases and is a smooth,
systematic misfit in the body of the density, not random jitter. That is what
makes it reducible by better conditioning and capacity, not only by more paths.

\subsection{Effect of the Input Features}
\label{sec:results_features}
The input features are the larger of the two data-side levers (the other being the
training set, Section~\ref{sec:results_data}), and they act directly on the hard
regions of Section~\ref{sec:results_map}. Adding the persistence coordinate
$\log(1.01-\kappa)$ roughly halves the test error against plain log-features, from
$3.3\times$ to $1.55\times$ the floor; the short-$T$ indicators give a further
small gain, from $1.55\times$ to $1.50\times$, concentrated in the short maturities
they target (Table~\ref{tab:features}).

\begin{table}[htbp]
\centering
\caption{Effect of the input feature transforms on test CDF error
  ($[256,256]$, $5\times10^{7}$ samples, single seed). Each row adds a feature to the
  one above: the persistence coordinate roughly halves the error; the short-$T$
  indicators give a further small gain at the maturities they target.}
\label{tab:features}
\begin{tabular}{l r r}
\toprule
Feature set & CDF error & $\times$ floor \\
\midrule
base (log features) & $4.2\times10^{-4}$ & $3.3$ \\
\quad + persistence coordinate $\log(1.01-\kappa)$ & $2.0\times10^{-4}$ & $1.55$ \\
\quad + short-$T$ indicators & $1.9\times10^{-4}$ & $1.50$ \\
\bottomrule
\end{tabular}
\end{table}

\subsection{Effect of Training-Set Size and Coverage}
\label{sec:results_data}
\begin{figure}[htbp]
  \centering
  \includegraphics[width=0.66\textwidth]{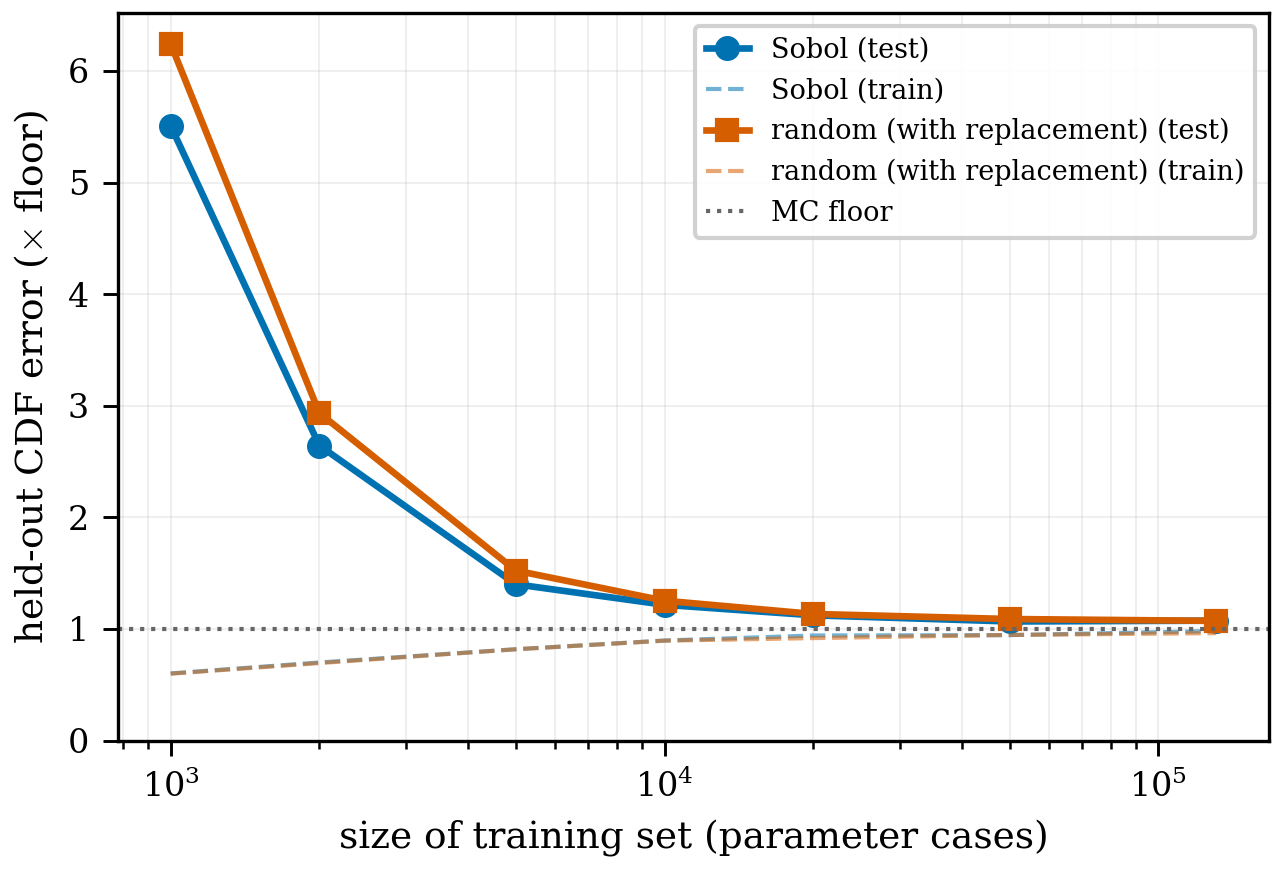}
  \caption{Data efficiency: test CDF error (solid) and training error
    (dashed), in floor units, versus the number of training parameter cases, under the
    best configuration (a $[512,256,128]$ network with the cosine schedule) at the
    converged $10^{8}$-sample budget. The Sobol design and an equally sized
    uniform-random subset (drawn with replacement) coincide at the full set, both
    reaching about $1.07\times$ the floor; Sobol leads only where cases are sparse
    (below about $10^{4}$). The spread between the dashed and solid curves is the
    train/test gap: the training error stays below the floor at every set size, deepest
    when cases are scarce, and the gap narrows to about $0.09\times$ at the full set.}
  \label{fig:data_efficiency}
\end{figure}

This experiment asks how many parameter cases the surrogate needs to see. With too few,
it has to interpolate across gaps in the parameter space and the test error suffers;
in the four-part error split of Section~\ref{sec:anatomy} this is the coverage term.
Figure~\ref{fig:data_efficiency} varies the number of training cases under the best
configuration. The test error drops steeply as cases are added, from about $5.5\times$
the floor at a thousand cases (Sobol) to $1.07\times$ at the full $1.3\times10^{5}$, the best
the method reaches (Table~\ref{tab:pricing_speed}). Most of the gain arrives by about
$10^{4}$ cases; the last stretch down to $1.07\times$ takes the remaining cases up to
$10^{5}$.

The training error moves the other way. It stays below the floor at every size, rising
from about $0.60\times$ at a thousand cases to $0.98\times$ at the full set. The network
fits the finite Monte Carlo labels a little more tightly than their own sampling noise,
most of all when cases are scarce, which is plain overfitting. As more cases are added
the training error climbs back toward the floor and the test error comes down to meet
it, leaving a small gap of about $0.09\times$ at the full set. So a shortage of cases is
what holds the error back below about $10^{4}$, and past $10^{5}$ there is little left to
gain.

Sobol sampling helps only when cases are scarce. Below about $10^{4}$ cases it beats an
equal-sized random subset (drawn with replacement) by up to roughly ten percent at a
thousand cases. The margin fades as the set grows, and the two match at the full set,
where either way of placing cases fills the space densely.

\subsection{Effect of the Training Objective}
\label{sec:results_loss}
\begin{figure}[t]
  \centering
  \includegraphics[width=\textwidth]{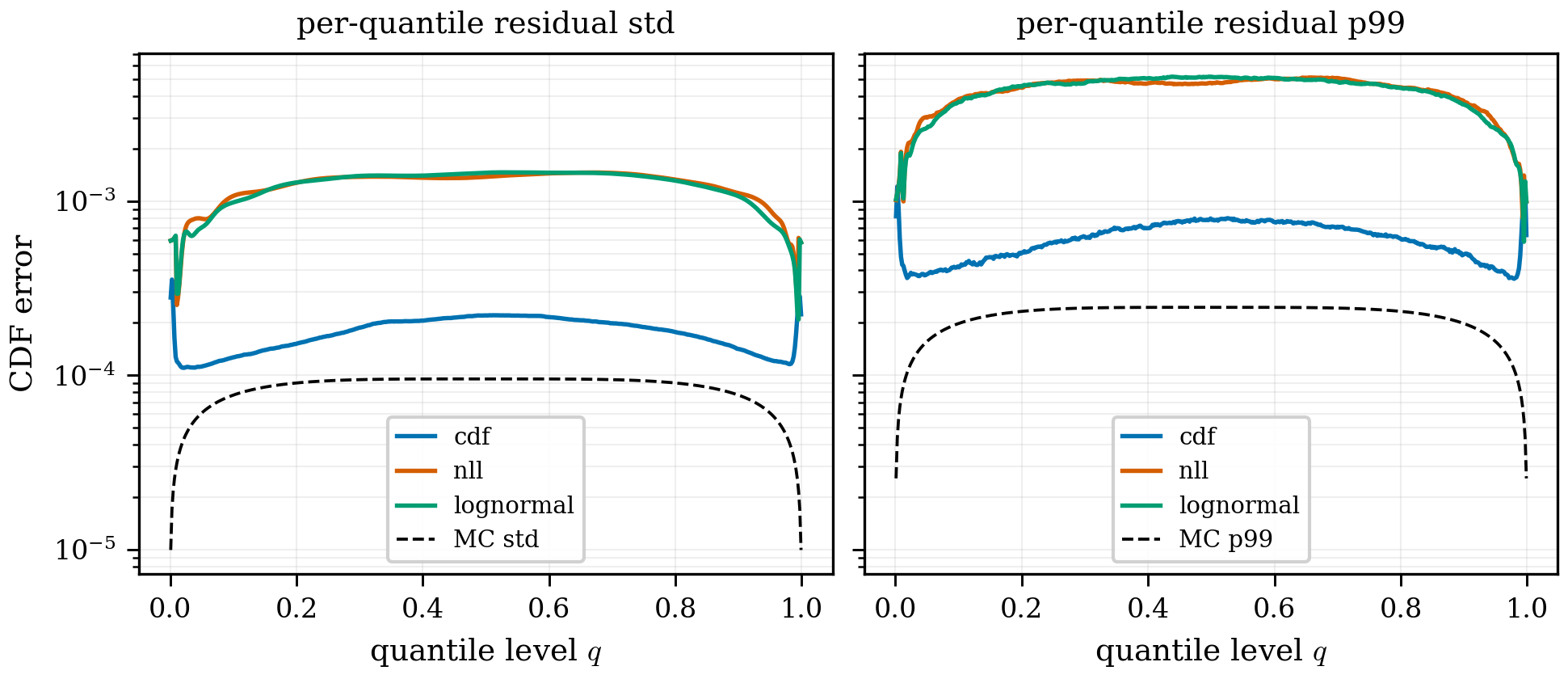}
  \caption{Per-quantile CDF residual for the three training losses, on a shared log
    axis: standard deviation (left, against the centered Monte Carlo std) and $99$th
    percentile of the absolute residual (right, against the Monte Carlo p99,
    $2.58\times$ the std). The CDF loss (blue) stays within a small multiple of the floor at every quantile;
    the likelihood losses (nll, lognormal) are several times higher throughout. The
    upturn of the CDF-loss curve at $q\to0,1$, where the floor instead decays, is the
    extreme-quantile spike discussed in the text.}
  \label{fig:losstail}
\end{figure}

The training objective decides which error the network actually minimises.
Section~\ref{sec:loss} argued that matching the CDF in quantile space is the right
choice; training identical networks under the three candidate objectives confirms
it. Figure~\ref{fig:losstail} compares their per-quantile residual, its standard
deviation and $99$th percentile across the test set. The CDF loss stays within a
small multiple of the floor at every quantile, in both panels, while the
likelihood objectives (nll, lognormal) are several times higher across the whole
range, tails included. Overall the gap is almost an order of magnitude: about
$1.1\times$ the floor for the CDF loss against ${\approx}8.5\times$ for the
likelihood objectives, on both the typical error and its $99$th percentile.

For the CDF loss, most of the residual matches the Monte Carlo noise in both size
and shape (the flat-topped $\sqrt{q(1-q)}$ profile), so across the body of the
distribution the error is sampling noise, not systematic misfit. The exception is
the extreme-quantile spike: as $q\to0,1$ the floor decays to zero, but the
residual instead rises into a narrow spike at the very edges, the largest-residual
quantiles already flagged in Section~\ref{sec:floor}. The spike is a capacity
limit of the finite mixture, not sampling noise: the Gaussian components have light
tails, whereas the true GJR--GARCH density inherits the heavier Student-$t$ tails
of its innovations, so the mixture cannot track the CDF through the last fraction
of a percent ($q\lesssim0.005$ and $q\gtrsim0.995$). It is small in absolute
terms, and a heavier-tailed component family does not remove it: replacing the
Gaussian components with Laplace ones (exponential tails, also closed-form to
price) leaves the spike unchanged and raises the overall test CDF error, to
$1.22\times$ the floor against $1.08\times$ for the Gaussian. We keep the Gaussian
mixture and treat the spike as a finite-mixture limitation.

\subsection{Effect of Model Capacity}
\label{sec:results_frontier}
\begin{figure}[t]
  \centering
  \includegraphics[width=\textwidth]{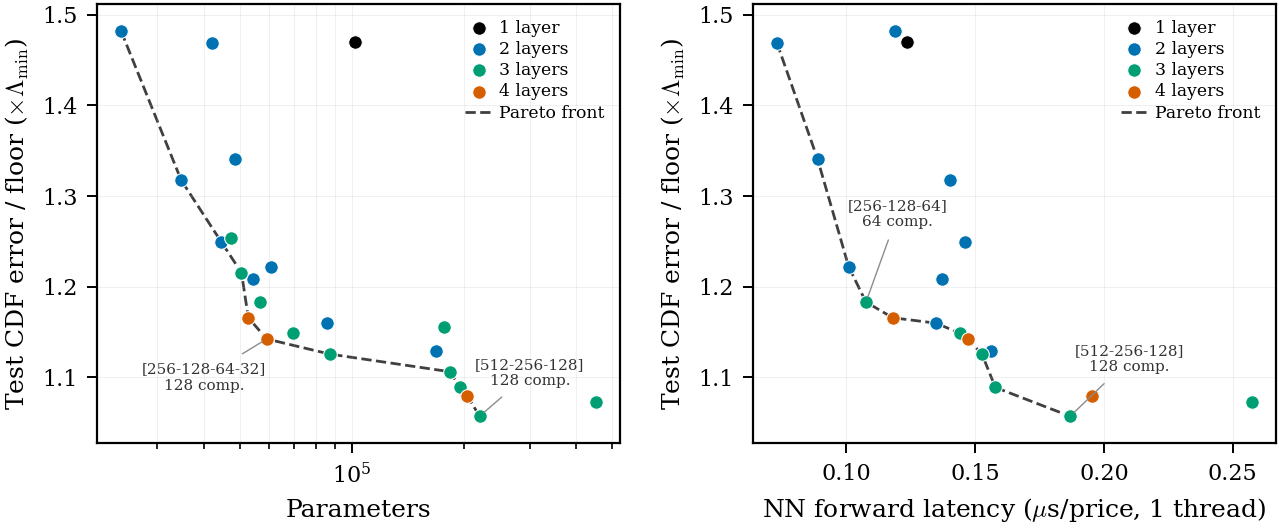}
  \caption{Capacity and speed--accuracy frontier at the converged $10^{8}$-sample
    budget (cosine schedule, CDF loss, warped-$T$ features, full training set; single
    seed). Left: test CDF error in floor units versus parameter count, coloured by
    network depth, with the Pareto front dashed. Right: the same error versus the
    network's forward latency per option price on a single CPU thread, which excludes
    the closed-form pricing step. Points span single-layer trunks, two-layer trunks,
    and three- and four-layer tapered funnels, with mixtures of $16$ to $128$
    components. Tapered funnels occupy the front at every size: the most accurate
    $[512,256,128]$ network ($\approx2.2\times10^{5}$ parameters) reaches $1.07\times$
    the floor, the small champion $[256,128,64]$ matches a far larger uniform
    $[256,256]$ trunk at under half the parameters, and a single wide layer is the
    least efficient use of capacity. The Monte Carlo floor is the horizontal
    reference.}
  \label{fig:pareto}
\end{figure}

Capacity is the last controllable term, and it trades accuracy against cost. We
map the trade-off by training the same recipe at the converged $10^{8}$-sample
budget across many architectures, from a single hidden layer to four-layer tapered
funnels with $16$ to $128$ components, and timing each. Figure~\ref{fig:pareto}
plots test CDF error against parameter count (left) and against forward latency per
price (right), with the floor marked.

Two patterns stand out. First, accuracy improves smoothly with capacity, then
flattens within about $7\%$ of the floor; beyond roughly $2\times10^{5}$
parameters, more size buys almost nothing. Second, shape matters at a fixed
parameter budget. Because the surrogate learns a whole family of densities, not a
single market surface, depth pays off here: a tapered funnel, narrowing toward the
output, consistently beats a uniform-width trunk of the same size, with the three-
and four-layer funnels on the Pareto front and a single wide layer the least
efficient.

The most accurate model, $[512,256,128]$ with $128$ components
($\approx2.2\times10^{5}$ parameters), reaches $1.07\times$ the floor. A
deeper-than-wide $[256,128,64]$ network matches a far larger uniform $[256,256]$
trunk at under half the parameters, a good default when size or memory is tight.

The two operating points of Table~\ref{tab:pricing_speed} separate the two stages
of pricing. The network forward pass is cheap and nearly flat across the frontier,
$0.07$ to $0.26\,\mu$s per price on a single CPU thread even for the largest
network, so the network is never the bottleneck. End-to-end cost is set by the
closed-form pricing step, which sums one Black term per mixture component and so
scales with the component count. On a single CPU thread this puts the
$128$-component model at $4.7\,\mu$s per price and the $64$-component model at
$0.9\,\mu$s; on a batched GPU both fall to about $0.36\,\mu$s. Sub-microsecond
pricing therefore holds for the small model on CPU and for either model on GPU,
while the most accurate model costs a few microseconds on a single CPU thread.

The two devices differ far more in throughput than in single-price latency, and
the reason is batch size. A single surface is a small batch, so the GPU's kernels
spend most of their time on fixed launch overhead, with too little work to fill the
cores; its per-price latency sits at a flat ${\approx}0.36\,\mu$s, only a little
below the single-thread CPU. Under bulk revaluation the batch is large: the launch
overhead spreads across many prices, the cores saturate, and the GPU reaches tens
of millions of prices per second, $30$ to $80\times$ the single-thread CPU, whose
throughput is essentially the reciprocal of its per-price latency: a serial loop
gains nothing from batching. In practice, a CPU prices one surface interactively at low
latency, and a GPU pays off for revaluing a whole book. Either way the surrogate is
orders of magnitude faster than a matched simulation, which we quantify next.

\begin{table}[htbp]
\centering
\caption{Two operating points from the frontier and their pricing cost on each device
  (CPU: single thread; GPU: batched; \texttt{float32}). \emph{Precision} is the test CDF
  error in floor units; latency is microseconds per option price and throughput is prices
  per second. The small model gives up little accuracy for roughly a quarter of the
  parameters and several times the CPU throughput; on GPU both price tens of millions of
  options per second at a flat ${\approx}0.36\,\mu$s each.}
\label{tab:pricing_speed}
\begin{tabular}{l c c c c c c}
\toprule
 & precision & & \multicolumn{2}{c}{CPU (1 thread)} & \multicolumn{2}{c}{GPU (batched)} \\
\cmidrule(lr){4-5}\cmidrule(lr){6-7}
Model & ($\times$floor) & params & $\mu$s/price & prices/s & $\mu$s/price & prices/s \\
\midrule
$[512,256,128]$, 128 comp.\ & $1.07$ & $2.2\times10^{5}$ & $4.7$ & $0.21$M & $0.36$ & $16$M \\
$[256,128,64]$, 64 comp.\   & $1.18$ & $5.7\times10^{4}$ & $0.9$ & $1.1$M  & $0.35$ & $32$M \\
\bottomrule
\end{tabular}
\end{table}

\subsection{Pricing Speed versus Monte Carlo Simulation}
\label{sec:results_speed}
Against Monte Carlo, the speedup is built into the method: the surrogate replaces
an $O(N\cdot\text{steps})$ simulated density with a single $O(1)$ forward pass. On
the same CPU, a matched-accuracy Monte Carlo price (the same $N=10^{7}$ paths over
$1000$ steps that set the training target) takes about $48$ s per option chain
($8$ maturities $\times$ $16$ strikes). The surrogate prices the same chain in
about $0.1$ ms end to end, at roughly $0.9\,\mu$s per price for the small model of
Table~\ref{tab:pricing_speed}: a speedup of order $4\times10^{5}$. The Monte Carlo
run used $6$ CPU threads against the surrogate's single thread, so the per-thread
gap is larger still.

Because the forward pass is also batched and differentiable, two things come
almost for free: a whole surface recomputes interactively, and parameter gradients
are available for calibration. Both are impractical with repeated simulation.

%% file: sections/13_pricing_behaviour.tex
\section{Pricing Behaviour and Sensitivities}
\label{sec:behaviour}
A usable surrogate must also produce \emph{economically sensible} surfaces, not
just hit the floor. The plots below vary one parameter at a time, at a fixed
reference volatility level, and read off the implied-volatility smile and the
return density at a representative maturity.

\begin{figure}[t]
  \centering
  \includegraphics[width=\textwidth]{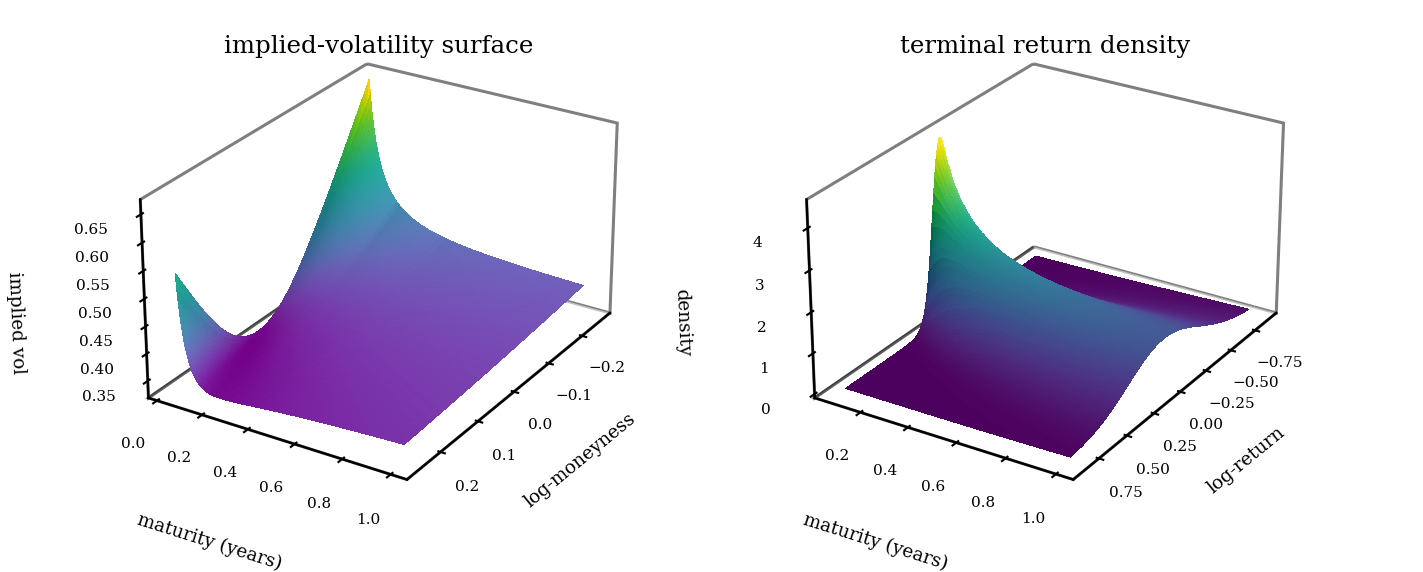}
  \caption{Surrogate output for the GJR--GARCH parameter set $\alpha=0.10$,
    $\gamma=0.30$, $\beta=0.55$, $\nu=5$, $\lambda=-0.20$, $(\sigma_0')^2=1$ at reference
    volatility $0.40$. Left: the implied-volatility surface across log-moneyness and
    maturity, with a negative skew that flattens as maturity grows. Right: the terminal
    log-return density across maturity, obtained from the same predicted mixture by one
    forward pass per maturity. Both surfaces follow from the GJR--GARCH dynamics alone.}
  \label{fig:iv_surface}
\end{figure}

\paragraph{The implied-volatility surface}
Figure~\ref{fig:iv_surface} shows two views of the surrogate's output for a
representative parameter set: the implied-volatility surface (left), across strikes
and maturities, and the terminal log-return density (right), across maturities. The
smile has the familiar shape of equity-index options, a negative skew that flattens
with maturity; the density is the matching left-skewed, heavy-tailed terminal
distribution, sharply peaked at short maturities and spreading as maturity grows.
Both shapes emerge purely from the GJR--GARCH dynamics, not from any fit to market
data.

\begin{figure}[t]
  \centering
  \includegraphics[width=\textwidth]{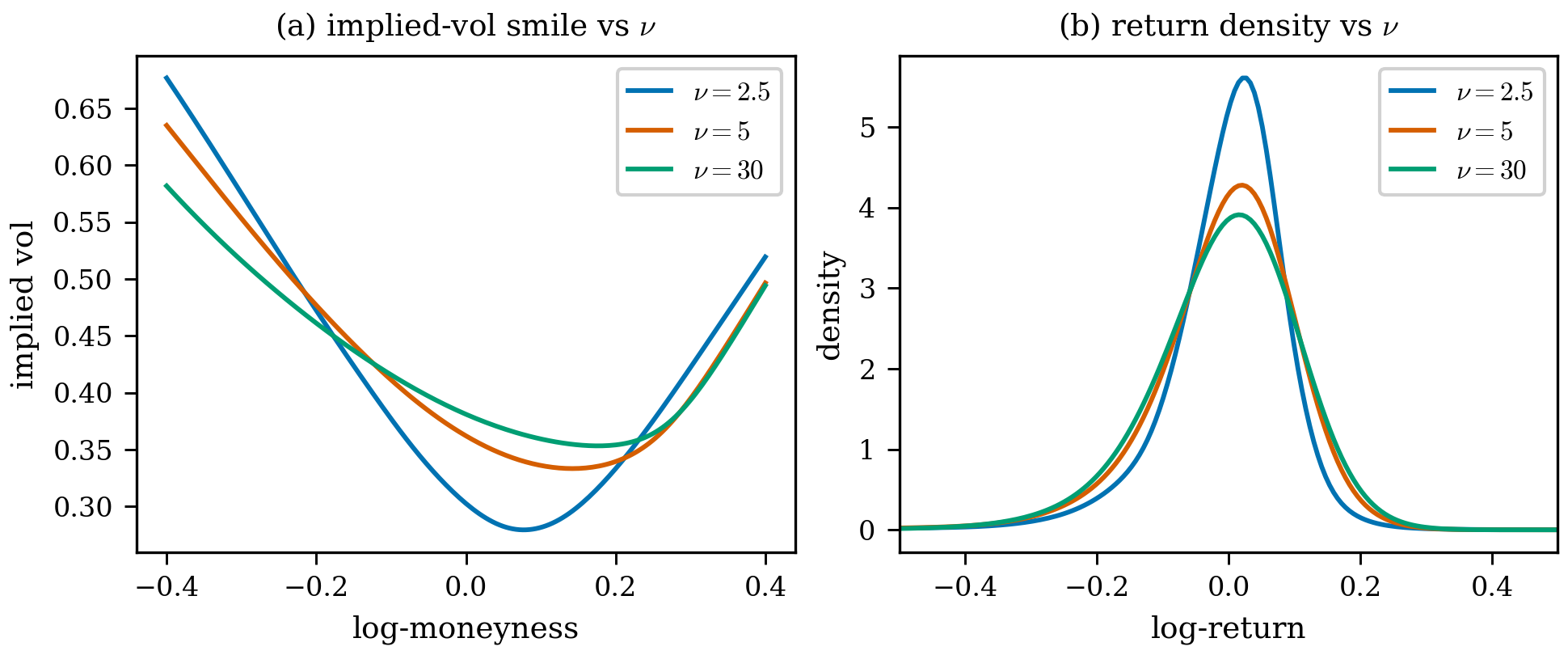}
  \caption{Effect of the tail parameter $\nu$ (skewed-$t$ degrees of freedom):
    lower $\nu$ thickens the tails of the return density and lifts the wings of
    the implied-volatility smile.}
  \label{fig:eta}
\end{figure}

\begin{figure}[t]
  \centering
  \includegraphics[width=\textwidth]{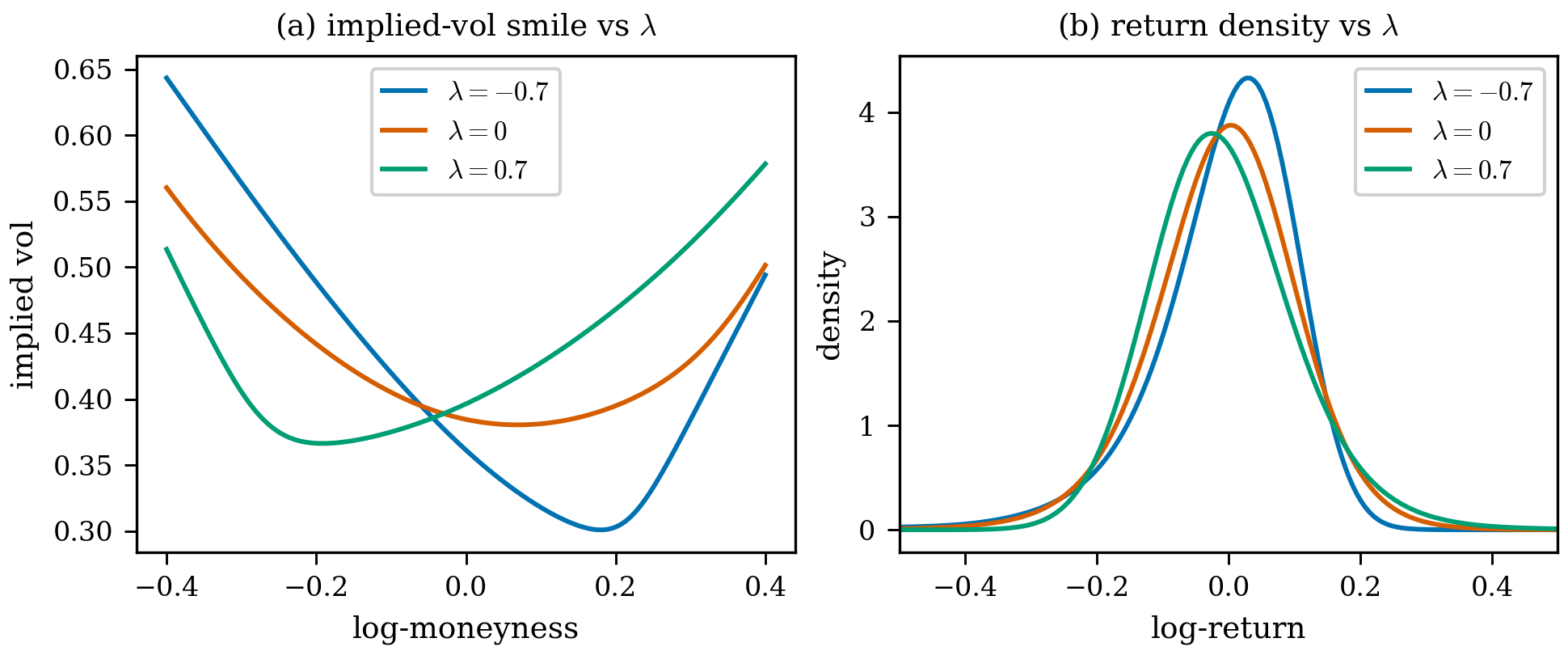}
  \caption{Effect of the skew parameter $\lambda$: it tilts the return density
    and sets the slope of the smile.}
  \label{fig:lam}
\end{figure}

\begin{figure}[t]
  \centering
  \includegraphics[width=\textwidth]{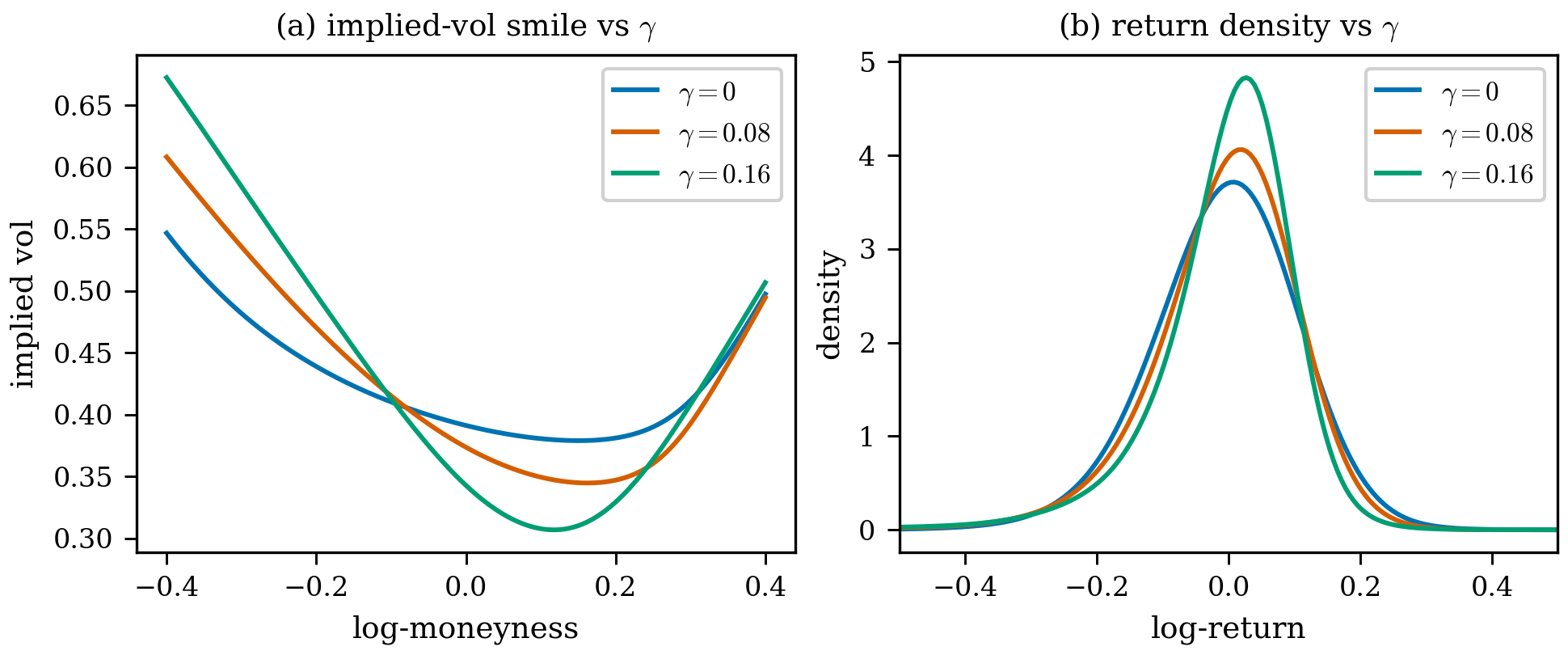}
  \caption{Effect of the leverage parameter $\gamma$: it amplifies volatility
    after negative shocks, steepening the downside of the smile.}
  \label{fig:gamma}
\end{figure}

\begin{figure}[t]
  \centering
  \includegraphics[width=\textwidth]{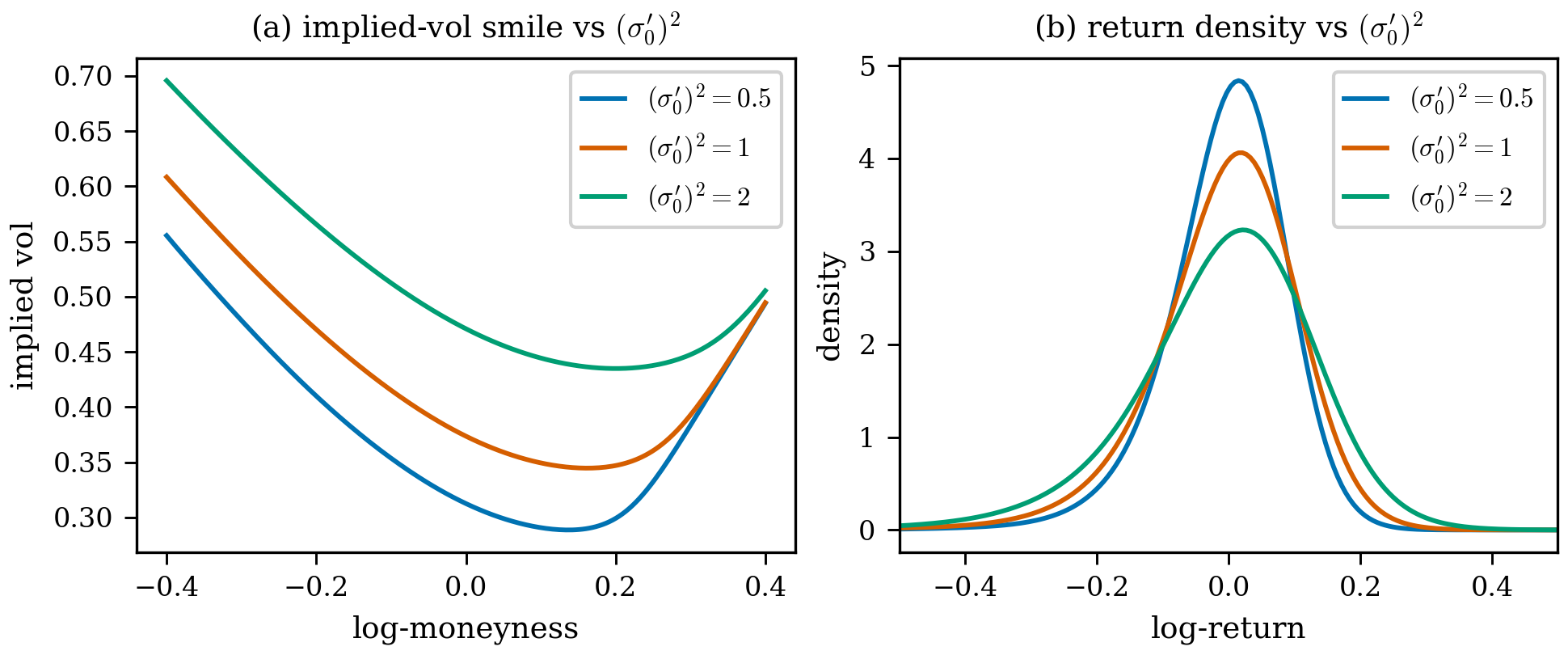}
  \caption{Effect of the initial variance $(\sigma_0')^2$: it sets the
    short-maturity level of the surface and decays as variance mean-reverts.}
  \label{fig:var0}
\end{figure}

\begin{figure}[t]
  \centering
  \includegraphics[width=\textwidth]{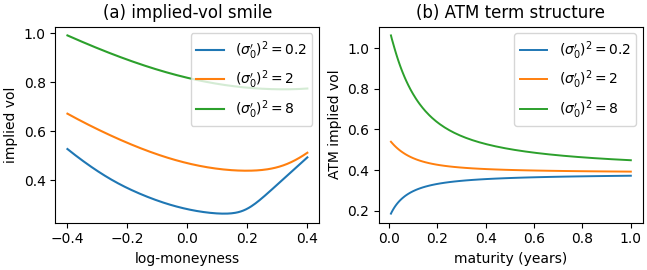}
  \caption{Effect of persistence: the coefficients $(\alpha,\gamma,\beta)$, via
    $\kappa=\alpha+\beta+\gamma\,p_-$, control how slowly a variance shock decays
    and hence the term structure of the smile.}
  \label{fig:var0_pers}
\end{figure}

\paragraph{Parameter sensitivities}
The reduced form isolates five shape controls:
\begin{itemize}[leftmargin=*,itemsep=0.25em]
  \item \textbf{Tails ($\nu$).} Here $\nu$ is the skewed-$t$ degrees of freedom
  (Section~\ref{sec:model}). Lower $\nu$ thickens both tails of the return density,
  lifting the wings of the smile relative to the at-the-money level. As $\nu$ grows
  the distribution approaches Gaussian and the smile flattens (Fig.~\ref{fig:eta}).
  \item \textbf{Skew ($\lambda$).} The skewed-$t$ asymmetry tilts the density and
  with it the slope of the smile; negative $\lambda$ gives the downward skew
  typical of equity indices (Fig.~\ref{fig:lam}).
  \item \textbf{Leverage ($\gamma$).} The asymmetric variance response amplifies
  volatility after negative shocks, steepening the downside of the smile and
  adding negative skew on top of the innovation skew (Fig.~\ref{fig:gamma}).
  \item \textbf{Initial variance $(\sigma_0')^2$.} The starting variance sets the
  short-maturity level of the surface; its influence decays as the variance
  mean-reverts towards the long-run level (Fig.~\ref{fig:var0}).
  \item \textbf{Persistence $(\alpha,\gamma,\beta)$.} Through
  $\kappa=\alpha+\beta+\gamma\,p_-$, the persistence coefficients govern how
  slowly an initial variance shock decays, controlling the term structure of the
  smile (Fig.~\ref{fig:var0_pers}).
\end{itemize}
These dependencies are smooth and move in the expected directions. That is both
economically sensible and the reason the surrogate interpolates so well between
training points.

%% file: sections/14_discussion.tex
\section{Discussion and Conclusion}
\label{sec:discussion}
We have presented a pretrained neural surrogate that prices European options
under the GJR--GARCH model without Monte Carlo simulation. The network learns the
terminal return density as a Gaussian mixture, from which prices, implied
volatilities and Greeks follow in closed form as weighted sums of Black formulas.
Three design choices make the surrogate transparent, not a black box: training in
a dimensionless reduced form removes the model's scale and shift ambiguities and
keeps the surrogate arbitrage-consistent; a CDF-matching loss aligns training with
pricing error and is justified by the bound~\eqref{eq:price_bound}; and a
closed-form, distribution-free Monte Carlo noise floor (Proposition~\ref{prop:floor})
ties the achievable accuracy to the simulation budget. The trained model reaches
that floor on an independent test set (CDF error $\approx1.4\times10^{-4}$, within
$10\%$ of the $N=10^{7}$ floor) while pricing an option chain about a factor of
$4\times10^{5}$ faster than matched-accuracy simulation
(Section~\ref{sec:results_speed}).

\paragraph{When a surrogate can replace simulation}
The noise floor gives a concrete criterion: a surrogate can stand in for the
Monte Carlo engine once its validation error comes within a small multiple of the
floor across the parameter region of interest, since near the floor the surrogate
and a fresh simulation are statistically comparable. Prices vary smoothly with the
parameters, which is what makes this attainable, and what lets the surrogate
interpolate between training points more accurately than any single noisy target.
Dense Sobol coverage of the parameter space and matching train/test errors are the
safeguards that make the claim verifiable rather than merely asserted.

\paragraph{Limitations}
The scope here is deliberately narrow: a single asset, the GJR--GARCH family, and
European payoffs. The surrogate prices payoffs that depend only on the terminal
return, and does not, as presented, handle path-dependent or early-exercise
products, which need the joint distribution of the whole path, not just the
terminal density. Accuracy is also bounded by the training budget and the coverage
of the sampled parameter region; extrapolation beyond it is not certified.
Finally, we validate against the model (the ``truth'' is GJR--GARCH itself); an
end-to-end market study (fitting GJR--GARCH to historical returns and comparing
surrogate prices to listed options) is a natural next step.

\paragraph{Outlook}
The same blueprint (a forward density operator trained to the noise floor in
dimensionless coordinates) extends directly to other GARCH variants, and with care
over discretisation error to continuous-time and rough volatility models. Other
directions: differentiable calibration that exploits the network's gradients,
joint training across model families, and multi-asset extensions. More broadly,
this is one instance of a wider pattern, using simulation-trained surrogates where
data are scarce but simulation is cheap, with one distinctive feature here: the
simulator's own sampling noise sets a principled, quantitative target for
sufficient accuracy.

%% file: sections/appendix_a_gjr_garch.tex
\section{GJR--GARCH: Innovations and Normalisation}
\label{app:gjr}
\subsection{Skewed Student-$t$ Innovations}
Following \citet{hansen1994autoregressive}, the skewed Student-$t$ distribution
$\skewt_{\nu,\lambda}(0,1)$ has degrees of freedom $\nu>2$ and skew parameter
$\lambda\in(-1,1)$. Define
\begin{equation}
  c = \frac{\Gamma\!\left(\tfrac{\nu+1}{2}\right)}
           {\sqrt{\pi(\nu-2)}\,\Gamma\!\left(\tfrac{\nu}{2}\right)},
  \qquad
  a = \frac{4\lambda c(\nu-2)}{\nu-1},
  \qquad
  b = \sqrt{1 + 3\lambda^2 - a^2}.
\end{equation}
With $z = bx + a$ and $x_0 = -a/b$, the density is
\begin{equation}
  f(x;\nu,\lambda) = \frac{c}{b}
  \begin{cases}
    \Big[\,1 + \dfrac{1}{\nu-2}\!\left(\dfrac{z}{1-\lambda}\right)^{2}\Big]^{-\frac{\nu+1}{2}},
      & x < x_0,\\[1.5ex]
    \Big[\,1 + \dfrac{1}{\nu-2}\!\left(\dfrac{z}{1+\lambda}\right)^{2}\Big]^{-\frac{\nu+1}{2}},
      & x \ge x_0.
  \end{cases}
\end{equation}
This parameterisation yields standardised innovations with $\E[z_t]=0$ and
$\operatorname{Var}(z_t)=1$, so that $z_t\sim\skewt_{\nu,\lambda}(0,1)$ enters the
recursion~\eqref{eq:gjr} directly through $\epsilon_t=\sigma_t z_t$.

\subsection{Stationarity and Shape-Only Normalisation}
The process is covariance-stationary when
\begin{equation}
  \kappa = \alpha + \beta + \gamma\,p_- < 1, \qquad
  p_- = \E\!\big[z_t^2\,\mathbf{1}\{z_t<0\}\big].
\end{equation}
Under the dimensionless reduced form used for training, the long-run variance is
normalised to one. With $v=\omega/(1-\kappa)$ the stationary value of
$\E[\sigma_t^2]$, the conditional variance
$\E[\sigma_t^2]=v+\kappa^{t}(\sigma_0^2-v)$ of Section~\ref{sec:model}
(eq.~\ref{eq:condvar_phys}) relaxes to $v$, and the reduced variable
$(\sigma_t')^2=\sigma_t^2/v$ has unit long-run variance. The learned mapping
therefore depends only on the dimensionless shape of the return distribution, not
on absolute scale or mean level. The one-sided second moment $p_-$ is evaluated by
numerical integration of $x^2 f(x)$ over $(-\infty,0)$, splitting the integral at
the kink $x_0=-a/b$ for accuracy.

%% file: sections/appendix_b_data_training.tex
\section{Data and Training Details}
\label{app:data}
\subsection{Dataset}
\begin{figure}[t]
  \centering
  \includegraphics[width=\textwidth]{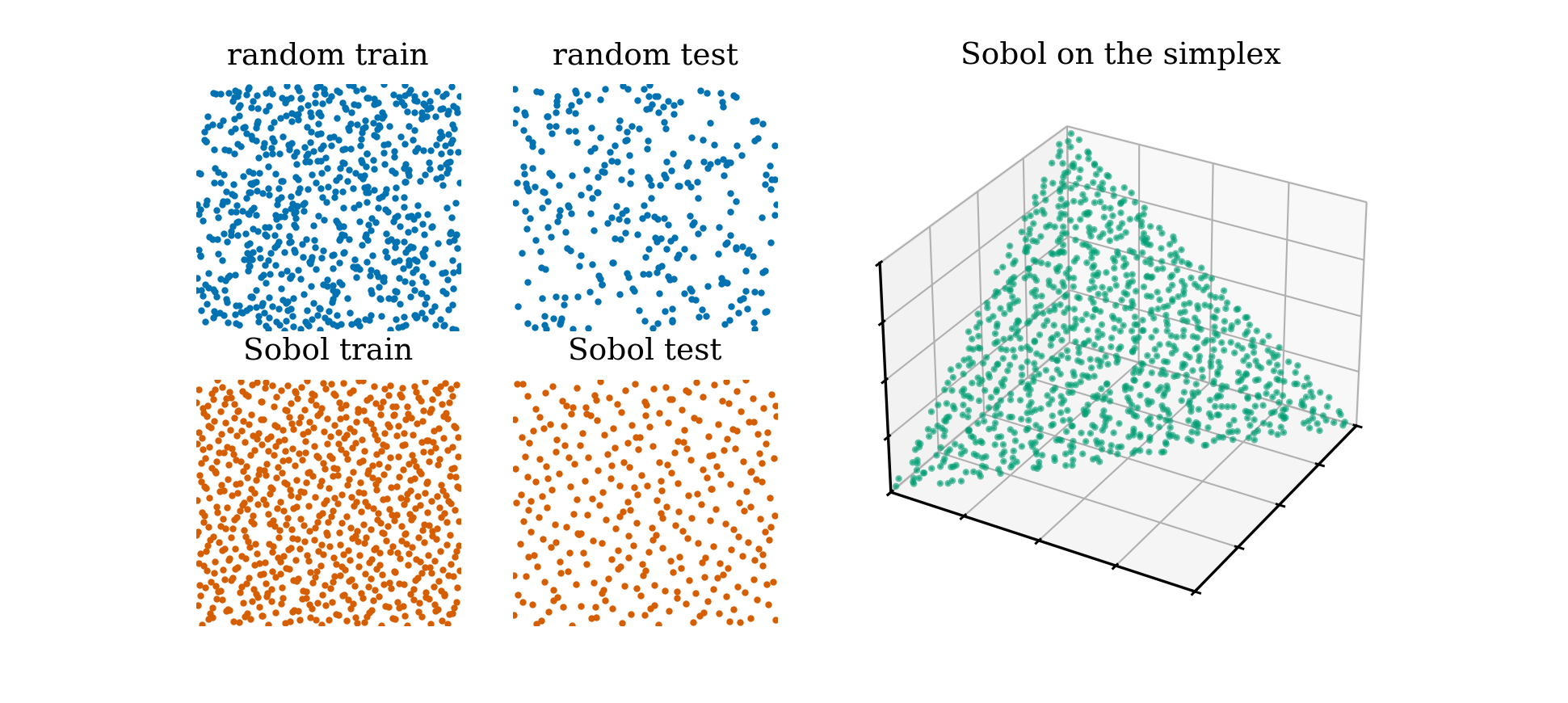}
  \caption{\textbf{Sobol vs.\ random sampling.}
    \textit{Top (blue):} random uniform samples show clustering and gaps.
    \textit{Bottom (red):} Sobol samples are evenly distributed.
    \textit{Right:} Sobol sampling within the simplex for the persistence
    coordinates $(\alpha,\gamma\,p_-,\beta)$ \citep{devroye1986}.}
  \label{fig:sobol}
\end{figure}

The training data is the public Hugging Face dataset
\texttt{simu-ai/garch\_densities}. Each example stores the reduced-form
parameters $(\alpha,\gamma,\beta,(\sigma_0')^2,\nu,\lambda)$, a maturity $T$ (in
steps), the $Q=512$ probability levels $p$ uniformly spaced in $[0.001,0.999]$,
and the corresponding standardised terminal-return quantiles $x$. The training
split contains $2^{17}\approx131$k parameter cases simulated with $N=10^{7}$
paths over $1{,}000$ timesteps; the test split contains $2^{15}\approx32$k
independent cases drawn from a disjoint Sobol stream. Figure~\ref{fig:sobol}
contrasts the even Sobol coverage with the clustering and gaps of uniform-random
sampling.
\subsection{Input Features}
The base network input is the eight-dimensional vector
\[
  \big(\alpha,\ \gamma,\ \beta,\ \log\!\big((\sigma_0')^2\big),\ \log\nu-1,\ \lambda,\
  \log(T-0.8)-1,\ \mathbf{1}\{T=1\}\big),
\]
i.e.\ the reduced-form parameters with the initial variance $(\sigma_0')^2$, the
degrees of freedom $\nu$ and the maturity $T$ log-transformed for better
conditioning, plus a single-step indicator $\mathbf{1}\{T=1\}$ for the shortest
horizon. The reported model adds three further features
(Section~\ref{sec:results_features}, Table~\ref{tab:features}): the persistence
coordinate $\log(1.01-\kappa)$, which stretches the near-nonstationary corner
$\kappa\to1$, and the short-horizon indicators $\mathbf{1}\{T=2\}$ and
$\mathbf{1}\{T=3\}$; together with the base $\mathbf{1}\{T=1\}$ these flag the
shortest horizons $T\in\{1,2,3\}$ of Section~\ref{sec:errors}, for an
eleven-dimensional input in total. The persistence coordinate uses a small floor
inside the logarithm ($1.01-\kappa\ge10^{-3}$) to remain finite at $\kappa\to1$.

\subsection{Optimisation}
\label{app:opt}
Training uses Adam on the CDF loss~\eqref{eq:cdf_loss} with minibatches of a few
hundred cases and the hold-then-decay learning-rate schedule of
Section~\ref{sec:errors} (cosine for the reported model): a high initial rate of
$2\times10^{-3}$ held to capture structure, then annealed to a final
$\eta^\star\approx6\times10^{-5}$, over a budget of $5\times10^{7}$ to $10^{8}$
presented cases. The architecture is one to four hidden layers of $32$ to $512$
units (plus one $768$-wide probe), as uniform trunks or tapered funnels, with a
smooth/gated activation (GELU by default) and $16$ to $128$ mixture components. An
optional hard-example ring buffer re-presents the worst-fitting cases to control
the maximum loss, and optional Brownian-bridge noise can be added to the targets
to study the smoothness of the learned density.

\subsection{Asset-Class Forwards}
The mixture pricer~\eqref{eq:mixture_price} is agnostic to the asset class, which
enters only through the choice of forward $F_0$ in the drift
correction~\eqref{eq:delta}: $F_0=F$ for futures (the quoted level);
$F_0=S_0 e^{(r-q)T}$ for stocks with dividend yield $q$;
$F_0=S_0 e^{(r_d-r_f)T}$ for FX with domestic/foreign rates $r_d,r_f$; and
$F_0=S_0 e^{(r+c-y)T}$ for commodities with carry $c$ and convenience yield $y$.

%% file: sections/appendix_c_notation.tex
\section{Notation}
\label{app:notation}
\begin{tabularx}{\textwidth}{@{}p{0.24\textwidth} Y@{}}
\toprule
\textbf{Symbol} & \textbf{Description} \\
\midrule
$r$ & Constant risk-free rate used for discounting. \\
$F_t$ & Forward price of the underlying; terminal value $F_T = F_0 e^{X_T}$. \\
$x_t$ & One-period log-return increment in the GJR--GARCH recursion. \\
$x_t'$ & Reduced one-period return $(x_t-\mu)/\sqrt{v}$ (zero mean, unit long-run
        variance). \\
$X_T$ & Standardised cumulative log-return over horizon $T$,
        $X_T=\tfrac{1}{s_T}\sum_{t=1}^{T} x_t'$, where $s_T$ is the exact cumulative
        standard deviation (equation~\ref{eq:cumvar}). \\
$p(x)$ & Density of the terminal return $X_T$ ($x$ a dummy variable). \\
$\Phi(\cdot)$ & Standard normal CDF (as in the Black formula). \\
$K,\ T$ & Option strike; time to maturity (years). \\
$C(K)$ & Present value of a European option with strike $K$, maturity $T$. \\
$g(F_T)$ & Terminal payoff (e.g.\ $\max(F_T-K,0)$ for a call). \\
$\sigma_t,\ \sigma_t^2$ & Conditional volatility and variance at time $t$. \\
$\epsilon_t=\sigma_t z_t$ & Return innovation with standardised shock $z_t$. \\
$z_t\sim\skewt_{\nu,\lambda}(0,1)$ & Hansen skewed Student-$t$ shock,
        $\E[z_t]=0$, $\E[z_t^2]=1$. \\
$\alpha,\gamma,\beta$ & GJR--GARCH reaction, leverage, persistence coefficients. \\
$\omega,\ \mu$ & Variance intercept; mean return. \\
$\nu,\ \lambda$ & Skewed-$t$ degrees of freedom ($\nu>2$) and skew
        ($\lambda\in(-1,1)$). \\
$p_-$ & $\E[z_t^2\,\mathbf{1}\{z_t<0\}]$, the downside variance share
        ($=\tfrac12$ if $\lambda=0$). \\
$\kappa$ & Persistence $\alpha+\beta+\gamma\,p_-$; stationarity requires
        $\kappa<1$. \\
$v$ & Long-run (stationary) variance $\omega/(1-\kappa)$; the conditional variance
        relaxes to it (equation~\ref{eq:condvar_phys}). \\
$(\sigma_0')^2$ & Reduced initial variance $\sigma_0^2/v$, a free coordinate of
        $\theta_{\DRF}$. \\
$\theta_{\DRF}$ & Reduced-form parameters $(\alpha,\gamma,\beta,(\sigma_0')^2,\nu,\lambda)$. \\
$\modelparams$ & Volatility-model parameter vector input to the network
        ($\equiv\theta_{\DRF}$). \\
$\netparams$ & Learnable weights of the neural network. \\
$M$ & Number of Gaussian mixture components. \\
$w_i,\mu_i,\sigma_i$ & Weight, mean, std.\ of the $i$th mixture component. \\
$\delta$ & Drift correction enforcing the forward, equation~\eqref{eq:delta}. \\
$Q$ & Number of quantile levels in the CDF loss ($Q=512$). \\
$q,\ x_q$ & Quantile level; corresponding return threshold $\cdfmc^{-1}(q)$. \\
$N$ & Monte Carlo paths per parameter case ($N=10^{7}$). \\
$\cdfmc(x)$ & Empirical CDF from $N$ Monte Carlo samples. \\
$\Lcdf,\ \Lmin$ & CDF-matching loss; Monte Carlo noise floor $\sqrt{1/(6N)}$. \\
$\varepsilon_{\text{test}}$ & Out-of-sample (test) CDF error, the accuracy metric
        reported throughout. \\
$e_{\mathrm{cap}},e_{\mathrm{cov}},e_{\mathrm{opt}}$ & Capacity, coverage, and
        optimiser contributions to the test error, adding in quadrature with the
        floor: $\E[\varepsilon_{\text{test}}^2]\approx\Lmin^2+e_{\mathrm{cap}}^2
        +e_{\mathrm{cov}}^2+e_{\mathrm{opt}}^2$ (equation~\ref{eq:anatomy}). \\
$\eta,\ \eta^\star$ & Learning rate; final rate
        $\approx(B/c_{\mathrm{opt}})\Lmin^2\approx6\times10^{-5}$. \\
$B$ & Minibatch size (a few hundred cases). \\
$c_{\mathrm{opt}}$ & Optimiser constant in
        $e_{\mathrm{opt}}\approx\sqrt{c_{\mathrm{opt}}\,\eta/B}$. \\
$\Normal(x\given\mu,\sigma^2)$ & Normal density with mean $\mu$, variance
        $\sigma^2$. \\
\bottomrule
\end{tabularx}